\documentclass[a4paper,fleqn,usenatbib,useAMS]{mnras}
\usepackage{graphicx}
\usepackage{amssymb}
\usepackage[T1]{fontenc}
\usepackage{ae,aecompl}
\usepackage{times}
\usepackage{multirow}

\def\JPhysAmt{J. Phys. A: Math. Theor.}
\def\EPJST{Eur. Phys. J. Spec. Top.}
\def\QJPAM{Quart. J. Pure Appl. Math.}
\def\CMDA{Celest. Mech. Dyn. Astron.}
\def\IJBC{Int. J. Bif. Chaos}
\def\LNP{Lect. Notes Phys.}

\title[Chaotic motion in a time-dependent barred galaxy]{Chaotic motion and the evolution of morphological components in a time-dependent model of a barred galaxy within a dark matter halo}

\author[Machado $\&$ Manos]{R.~E.~G. Machado$^{\mathrm{1,2}}$\thanks{E-mail: rubens.machado@iag.usp.br} and T.~Manos$^{\mathrm{3,4,5}}$\\
$^1$ Departamento de Ciencias F\'isicas, Universidad Andr\'es Bello, Av. Rep\'ublica 220, Santiago, Chile.\\
$^2$ Instituto de Astronomia, Geof\'isica e Ci\^encias Atmosf\'ericas, Universidade de S\~ao Paulo, R. do Mat\~ao 1226, 05508-090 S\~ao Paulo, Brazil.\\
$^3$ CAMTP - Center for Applied Mathematics and Theoretical Physics, University of Maribor, Krekova 2, SI-2000 Maribor, Slovenia.\\
$^4$ School of Applied Sciences, University of Nova Gorica, Vipavska 11c, SI-5270 Ajdov\v s\v cina, Slovenia.\\
$^5$ Institute of Neuroscience and Medicine, Neuromodulation (INM-7), Research Center J\"ulich, J\"ulich, Germany.
}

\date{Accepted 2016 March 7. Received 2016 March 7; in original form 2015 December 27}

\pubyear{2016}

\begin{document}
\label{firstpage}
\pagerange{\pageref{firstpage}--\pageref{lastpage}}
\maketitle

\begin{abstract}
% context
Studies of dynamical stability (chaotic versus regular motion) in galactic dynamics often rely on static analytical models of the total gravitational potential. Potentials based upon self-consistent $N$-body simulations offer more realistic models, fully incorporating the time-dependent nature of the systems.
% aims
Here we aim at analysing the fractions of chaotic motion within different morphological components of the galaxy. We wish to investigate how the presence of chaotic orbits evolves with time, and how their spatial distribution is associated with morphological features of the galaxy.
% methods
We employ a time-dependent analytical potential model that was derived from an $N$-body simulation of a strongly barred galaxy. With this analytical potential we may follow the dynamical evolution of ensembles of orbits. Using the Generalized Alignment Index (GALI) chaos detection method, we study the fraction of chaotic orbits, sampling the dynamics of both the stellar disc and of the dark matter halo.
% results
Within the stellar disc, the global trend is for chaotic motion to decrease in time, specially in the region of the bar. We scrutinized the different changes of regime during the evolution (orbits that are permanently chaotic, permanently regular, those that begin regular and end chaotic, and those that begin chaotic and end regular), tracing the types of orbits back to their common origins. Within the dark matter halo, chaotic motion also decreases globally in time. The inner halo ($r<5$ kpc) is where most chaotic orbits are found and it is the only region where chaotic orbits outnumber regular orbits, in the early evolution.
\end{abstract}


%%%%%%%%%%%%%%%%%%%%%%%%%%%%%%%%%%%%%%%%%%%%%%%%%%%%%
\section{Introduction} \label{intro}

% orbits in galactic dynamics
Orbits are generally regarded as the backbone of structure in galaxies. Exploring orbital properties in general -- and in particular the evolution of their dynamical stability -- is a fundamental aspect in improving our understanding of galactic structures as a whole. Our ability to explore the details of orbital stability in galaxies depends considerably on the adequacy of analytical models, which can be time-independent (TI) or time-dependent (TD). (See e.g. \citealt{Vasiliev2015} for a versatile method of creating self-consistent equilibrium models of galaxies using Schwarzschild orbit superposition). Studying the stability and the phase space structure via analytical models \citep[see e.g.][]{ManAthMNRAS2011} has proven to be quite useful \cite[for a review, see][]{ContBook2002}, as long as those potentials are realistic, in the sense of adequately representing the density distributions of real galaxies.

% chaos
An important aspect of the role of chaos in galactic dynamics is manifested in the form of so-called `stickiness' or weak chaos \citep[see e.g.][]{Kandrupetal2000,TerKand2004MNRAS,ConHar2010CeMDA}, a regime in which an orbit may not fill phase space as thoroughly as in the strongly chaotic case. As a result, orbits may spend significant periods in confined regimes, thus contributing to the rise of stable structures, rather than hindering it \citep[see e.g.][]{AthaRomMas2009MNRAS,AthaRomBosMas2010MNRAS,AthRomBosMas2009MNRAS, HarKalMNRAS2009,HarKalCont2011MNRAS,HarKalCont2011IJBC,ConHar2013MNRAS, KauCont1996A&A,PatAthQui1997ApJ,Pat2006MNRAS,RomMasAthGar2006A&A,RomAthMasGar2007A&A, TsoKalEftCon2009A&A,BruChiPfe2011A&A,BouManAntCeMDA2012}.

% chaos detection
Lyapunov exponents \citep[see e.g.][]{SkoLNP2010} have been extensively used for the detection of chaotic motion in several different models. However, there are often disadvantages which hinder their use. Several approaches exist to detect and quantify chaos, whose differences and efficacies have been thoroughly compared and discussed in the recent literature \citep[see][and references therein]{ContBook2002,MafDarCinGio2011CeMDA,MafDarCinGio2013MNRAS}. In \citet{LNinP2016} the reader may find a special volume with a more complete and recent review of the several chaos detection methods broadly used and their predictability as well as all the relevant references regarding their theoretical background, numerical implementation and applications in various models. \cite{Weinberg2015b} recently studied chaotic orbits in a time-independent model of a barred galaxy applying a new chaos detection method based on Kolmogorov-Arnold-Moser theory \citep{Weinberg2015a}. We use the Generalized Alignment Index (GALI) method \citep{ManBouSkoJPhA2013,SkoManLNP2016}.

% dark matter halo
Although orbital and dynamical analyses almost always focus on the stellar disc, detailed studies of the orbital structure of haloes have also been carried out \citep[e.g.][]{Valluri2010,Valluri2012}, characterizing the orbital families and investigating the effects that baryons play on the dark matter halo. Specifically, \cite{Valluri2013} studied the orbital structure of the stellar halo and also of the dark matter halo, and found that tidal debris in cosmological hydrodynamical simulations experience more chaotic evolution than in collisionless simulations. \cite{Price-Whelan2016} recently used a static triaxial potential (not containing baryons) to represent the Milky Way and study the presence of chaos in stellar streams resulting from tidal debris. Also with an analytical static triaxial model, \cite{Maffione2015} analysed the chaos onset time of particles in the stellar halo.

%  nbody simulation
In this paper, we will employ the time-dependent analytical model previously developed in \cite{ManosMachado2014}. It was derived from an $N$-body simulation of a disc galaxy within a live halo (i.e. both the stellar particles and the halo particles were responsive to their self-consistent potential). In that simulation \citep{MachadoAthanassoula2010}, interaction between disc and halo leads to the formation of a strong bar. The measured parameters were used as input to the analytical model, which was composed of three components: bar, disc and halo. Among several simplifying assumptions, we point out for example that the analytical bar is described  as a simple ellipsoid. While sufficient for most purposes (such as measuring quantities within the bar region), an ellipsoid is not a faithful representation of the actual shape of an evolving $N$-body bar. Nevertheless, the global dynamics of the system were quite well reproduced, as indicated by the comparison of the rotation curves of the $N$-body simulation and of the analytical model.

% this paper
Here, we extend our previous study, focusing on a detailed exploration of the fractions of chaotic motion both in the disc and in the halo. We aim to explore not merely the time evolution, but also the spatial distribution associated with different morphological components of the galaxy -- namely the bar, the ring, the intermediate region between the bar and the ring, and the outer disc; and in the case of the halo, the inner versus outer parts. Let us also stress the much richer orbital variety that the TD potential carries compared to the rather simpler TI analytical potentials, or those derived from snapshots of $N$-body simulations. I.e. orbits that evolve with such TD models can alternate both their spatial morphology and their stability (chaotic or regular) in time. This emanates from the change of stability of the main (but not only) families of stable (or unstable) periodic orbits, as the main parameters of the potential evolve simultaneously in a very complex manner in general. This drastically affects the stability of their nearby phase space regimes and the global stability in general (see e.g. \citealt{ManBouSkoJPhA2013,ManosMachado2014}) as well as orbital shape transformation from one type to another (e.g. a disc-like trajectory can change to a bar-like or a ring-like, etc). Both the changes of orbital shape and stability are features that take place in a typical $N$-body simulation and hence it can be very helpful to be able to study and understand these morphological and dynamical transitions and their effect in the global and local dynamics via such a simpler TD. Furthermore, once we have characterized the spatially resolved fractions of chaotic motion at given times, we can investigate the transitions between different regimes of stability. Finally, we may use this information to trace orbits back and uncover their common origins.

% sections
This paper is organized as follows. In Section~\ref{sec:techniques}, we summarize how an analytical potential model was derived from an $N$-body simulation, and we review the techniques of chaos detection which we will be using for our studies here. We present the results for the disc and for the halo in Sections~\ref{sec:disc} and~\ref{sec:halo} respectively, in which we explore the evolution of the fractions of chaotic orbits and their spatial distribution in time. In Section~\ref{sec:conclusion} we discuss and summarize our findings.

%%%%%%%%%%%%%%%%%%%%%%%%%%%%%%%%%%%%%%%%%%%%%%%%%%%%%
\section{Model and techniques} \label{sec:techniques}

Before starting the description of the TD model and chaos detection techniques, let us here mention that regarding this section we have a twofold aim: (i) to keep this article as self-contained as possible by providing all the basic information and properties for the reader to follow the upcoming results in the next sections and (ii) to avoid redundant repetition of all the details of the model/chaos detection methods which can be found in \citep{ManosMachado2014} for someone more interested in how the model was conceived and derived at first as well as more details and a short review on chaos detection methods in general. Hence, in the next subsections we try to give a concise description of the model and the tools used for the distinction between different dynamical states (regular from chaotic) of a given orbit.

\subsection{Analytical model from an $N$-body simulation}

A time-dependent analytical model was developed by \cite{ManosMachado2014} to represent the gravitational potential of a barred galaxy. Here we briefly summarize the main features of that model and the reader is referred to that paper for further details.

% n-body
To produce an astrophysically well-motivated model, we based it on one of the simulations described in \cite{MachadoAthanassoula2010}. We considered an $N$-body simulation of a disc galaxy embedded in a live spherical dark matter halo. The mass of the disc was $M_{d}=5\times 10^{10}~{\rm M}_{\sun}$, with an exponential density profile having radial scale length $R_{d}=3.5$~kpc and vertical scale height $z_{0}=0.7$~kpc. The dark matter halo had a \cite{Hernquist1993} density profile and was five times as massive as the disc. That simulation was a typically representative collisionless simulation of a strongly barred galaxy and it had been performed with 1.2 million equal-mass particles and carried out for approximately one Hubble time.

% analytical potential
Based on this $N$-body simulation, we then constructed an analytical model whose total gravitational potential was given by the sum of the potentials of the disc, bar and halo as \mbox{$V = V_{D}(t) + V_{B}(t) + V_{H}(t)$}. All these components were time-dependent, with parameters evolving in accordance with the behaviours measured from the simulation. Each individual component is represented as follows:

\begin{enumerate}

\item[\bf (a)] The disc is expressed by a Miyamoto-Nagai potential \citep{MNPASJ1975}:
\begin{equation}\label{eq:MNPot}
   V_{D}(t)=- \frac{GM_{D}(t)}{\sqrt{x^{2}+y^{2}+(A+\sqrt{z^{2}+B^{2}})^{2}}},
\end{equation}
where $A$ and $B$ are time-dependent parameters and describe its horizontal and vertical scale-lengths while $M_{D}(t)$ is the mass of the disc. Note that here `disc mass', does not include the bar, i.e. it only refers to the stellar component described by this axisymmetric potential.

\item[\bf(b)] A triaxial Ferrers bar \citep{Fer1877}, whose density is given by:
\begin{eqnarray}
  \rho(x,y,z) = \left\{
  \begin{array}{l l}
    \rho_{c}(1-m^{2})^{2}& \quad \textrm{if} \quad m<1,\\
    \quad 0& \quad \textrm{if} \quad m \geq 1,\\
  \end{array} \right.
\end{eqnarray}
where $\rho_{c}=\frac{105}{32\pi}\frac{G M_{B}(t)}{abc}$ is the central density, $M_{B}(t)$ is the mass of the bar, which changes in time, and $m^{2}=\frac{x^{2}}{a^{2}}+\frac{y^{2}}{b^{2}}+\frac{z^{2}}{c^{2}}$, \mbox{$a>b>c> 0$}, with $a,b$ and $c$ being the semi-axes of the ellipsoidal bar. The corresponding bar potential is:
\begin{equation}\label{Ferr_pot}
    V_{B}(t)= -\pi Gabc \frac{\rho_{c}}{3}\int_{\lambda}^{\infty}
    \frac{du}{\Delta (u)} (1-m^{2}(u))^{3},
\end{equation}
where $G$ is the gravitational constant (set to unity), $m^{2}(u)=\frac{x^{2}}{a^{2}+u}+\frac{y^{2}}{b^{2}+u}+\frac{z^{2}}{c^{2}+u}$, $\Delta^{2} (u)=({a^{2}+u})({b^{2}+u})({c^{2}+u})$, and $\lambda$ is the unique positive solution of $m^{2}(\lambda)=1$, outside of the bar ($m \geq 1$), while $\lambda=0$ inside the bar. The analytical expression of the corresponding forces are given in \cite{PfeA&A1984a}. In our model, the shape parameters (i.e. the lengths of the ellipsoid axes $a$, $b$ and $c$ are) are also functions of time. By construction, the bar grows with time in the TD analytical potential, as it does in the $N$-body simulation. Hence, the bar mass increases at the expense of the remainder of the disc mass. However, the total stellar mass remains always constant: $M_{B}(t)+M_{D}(t) = 5\times 10^{10}~{\rm M}_{\sun}$.

\item[\bf (c)] The spherical dark matter halo is given by a Dehnen potential \citep{DehnenMNRAS1993}:
\begin{equation}\label{eq:DehnenPot}
V_{H}(t)=\frac{GM_{H}}{a_H} \times\left\{
                      \begin{array}{lll}
                        -\frac{1}{2-\gamma}\left[1-\left(\frac{r}{r+a_H}\right)^{2-\gamma}\right]&, & \hbox{$\gamma \neq 2$,} \\
                        \ln \frac{r}{r+a_H}&, & \hbox{$\gamma = 2$.}
                      \end{array}
                    \right.
\end{equation}
where $M_{H}$ is the halo mass, $a_{H}$ is its scale radius and $\gamma$ (within $ 0 \leq \gamma < 3$) is a dimensionless parameter related to the inner slope. Contrary to the masses of disc and bar, the halo mass is considered constant in time. Nevertheless, the parameters $a_{H}$ and $\gamma$ are time-dependent. For $\gamma <2$ its finite central value is equal to $(2-\gamma)^{-1}GM_H/a_H$.
\end{enumerate}

We then measured from the $N$-body simulation the following parameters as a function of time: halo scale length, halo inner slope, disc vertical scale length, disc horizontal scale length, bar major axis, bar intermediate axis, bar minor axis, bar mass and bar pattern speed. Then we made fits to the time evolution of each of these parameters and supplied those results into the analytical model.

To provide the reader with some quantitative details about the model, here we give the approximate ranges of variation of the aforementioned parameters, as they evolve since $t=0$ until $t=12$ Gyr. The scale radius of the halo, $a_{H}$, varies in the approximate range from 3 to 6 kpc, while $\gamma$, the inner slope of the halo, varies roughly from 0 to 1. The horizontal and vertical scale lengths of the disc, $A$ and $B$, respectively go from 2.5 to 0.5 kpc, and from nearly 0 to 0.5 kpc. The bar mass grows from 0 to $3.3 \times 10^{10} {\rm M}_{\sun}$. Since the stellar mass is constant, the remainder of the disc mass consequently decreases by this same amount. The shape parameters of the bar, $a, b$ and $c$, start at nearly 0 and reach as much as 8, 2.8 and 1.9 kpc respectively. We acknowledge that the actual shape of the $N$-body bar is more complicated and thus ellipsoidal fits cannot be guaranteed to give very good approximations at all times. This is a convenient approach inasmuch as it allows the use of a triaxial Ferrers bar model. Furthermore, the mass encompassed by the ellipsoid gives an acceptable estimate of the mass of the bar, and the orientation of the major axis is quite well defined. Finally, as the bar grows stronger, its pattern speed $\Omega_b$ decreases greatly, from more than 70 to nearly 10 km s$^{-1}$ kpc$^{-1}$. All of these parameters -- but particularly the bar mass, bar length and bar pattern speed -- undergo their most important changes during the first $\sim$2.5 Gyr of the evolution.

With the above TD potential, we then construct a 3-degree-of-freedom (d.o.f.) Hamiltonian function which governs the motion of a star in a 3-dimensional rotating barred galaxy:
\begin{equation}\label{eq:Hamilton}
  H=\frac{1}{2} (p_{x}^{2}+p_{y}^{2}+p_{z}^{2})+ V(x,y,z,t) -
  \Omega_{b}(t) (xp_{y}-yp_{x}).
\end{equation}
The $x$ and $y$ refer to the directions along the major axis and intermediate axis of the bar respectively. The bar rotates around its short $z$-axis. The canonically conjugate momenta are expressed as $p_{x}$, $p_{y}$ and $p_{z}$, while $V$ is the total TD potential, $\Omega_b(t)$ represents the pattern speed of the bar and $H$ is the total energy of the orbit in the rotating frame of reference\footnote{That would be equivalent to the Jacobi constant for a TI Hamiltonian function.}.

% adequacy
This procedure involved numerous idealized simplifications, such as approximating the shape of the bar by an ellipsoid, etc. Nevertheless, these techniques worked in favour of the desired analytical simplicity, and generated a model that was able to reproduce several features with excellent agreement. For example, we found that the rotation curves were well recovered by the analytical model, indicating the adequacy of the global dynamics. Furthermore, the study of ensembles of orbits indicated that even morphological details were quite well reproduced. More details and explanations regarding the above TD potential and its parameters can be found in \cite{ManosMachado2014}.

\subsection{Techniques for detecting and measuring amount of chaos}

Let us here, for the sake of completeness, briefly recall how the two main chaos detection methods used throughout the paper, namely the Maximal Lyapunov Exponent (MLE) and the GALI method, are defined, calculated and more precisely which numerical procedures we are using (mainly for the GALI method for the goals of this work).

Considering the Hamiltonian function (equation~\ref{eq:Hamilton}), we derive the corresponding equations of motion together with the variational equations (see \citealt{ManosMachado2014} for more details). The latter ones govern the evolution of one or more deviation vectors $\mathbf{w}=(\delta x,\delta y,\delta z,\delta p_{x},\delta p_{y},\delta p_{z})$. The time evolution of such vectors constitutes the basic ingredient for the calculation of the MLE and the GALI chaos detection methods. For this purpose, one has to (numerically) solve simultaneously both the equations of motion and the variational equations (providing the time evolution of the orbit and of the deviation vectors respectively).

\subsubsection*{Lyapunov exponents}

For the computation of the MLE we follow the formulae and recipe proposed in \cite{BenGalStr1976PRA,ConGalGio1978PRA,Ben1980Mecc} and we define $\lambda_1$ as (see \citealt{SkoLNP2010} for a more recent description):
\begin{equation}
\label{LE}
\lambda_1 =\lim_{t \rightarrow \infty} \sigma_1(t),
\end{equation}
where:
\begin{equation}
\label{sigma_1}
\sigma_1(t)= \frac{1}{t}
\ln \frac{\|\mathbf{w}(t)\|}{\|\mathbf{w}(0)\|},
\end{equation}
is the so-called `finite-time MLE', with $\|\mathbf{w}(0)\|$ and $\|\mathbf{w}(t)\|$ representing the Euclidean norm of the deviation vector at times $t=0$ and $t>0$ respectively. In general, $\sigma_1(t)$ tends to zero (following a power law $\propto t^{-1}$) when the motion of an orbit is regular, and it converges to a non-zero value when the motion is chaotic. Let us stress here that in the case of conservative (TI) systems things are to some degree well distinguishable, i.e. orbits can be either periodic (stable or unstable) or regular or chaotic (see e.g. \citealt{LichLieb}) and hence the evolution and final value of the MLE can be well associated with the true nature of the orbit. Moreover, there is a further classification among the chaotic motion which has to do with the `degree' of dispersion or the time scale of its manifestation with respect to the system time scale. Therefore, in the literature one may find many studies on the so-called `weak or sticky' compared to `strongly' chaotic motion. The former is characterized typically by a smaller relatively positive MLE and confined (to some extent and/or for a certain time interval) diffusion in the configuration and/or phase space than the latter. In our previous article \citep{ManosMachado2014} we gave a brief overview of the recent literature on this particular kinds of motion emerging in conservative (TI) systems.

In this work, however, we consider a TD potential, a fact that gives rise to richer dynamics and behaviour for our orbits governed by it. The orbits can alternate (but not necessarily) their current dynamical state, from chaotic (or regular) to regular (or chaotic) over several time intervals of their evolution. As described more thoroughly in \cite{ManBouSkoJPhA2013} and also later in \cite{ManosMachado2014}, in such a case, the MLE (equation~\ref{LE}) cannot be used to safely characterize the asymptotic behavior of an orbit due to strong fluctuations caused be the dynamical transitions which take place as the potential evolves in time. Nevertheless, we may show sometimes the MLE for a sample of orbits. The main reason behind this is to get a more wide overview of the several dynamical transition taking place and examined by the main chaos detection tool we will be using, i.e the General Alignment Index (GALI, introduced in \citealt{SkoBouAntPhyD2007}).

\subsubsection*{The General Alignment Index (GALI)}

\noindent
In this study, we use the GALI method of chaos detection in the same manner as in our previous work \citep{ManosMachado2014}. For the calculation of the GALI index of order $k$ (GALI$_k$) one has to follow the evolution of $2 \leq k \leq N$ initially linearly independent deviation vectors $\textbf{w}_i(0)$, $i = 1,2,\ldots,k$, where $N$ denotes the dimensionality of the systems's phase space. GALI$_k$ is then defined \citep{SkoBouAntPhyD2007} as the volume of the $k$-parallelogram having as edges the $k$ unit deviation vectors $\hat{\textbf{w}}_i(t)=\textbf{w}_i(t)/\|\textbf{w}_i(t)\|$, $i = 1,2,...,k$. This volume can be expressed as the norm of the wedge product (denoted by $\wedge$) of these vectors:
\begin{equation}\label{GALI:0}
  \textrm{GALI}_{k}(t)=\parallel \hat{\textbf{w}}_{1}(t) \wedge \hat{\textbf{w}}_{2}(t) \wedge \ldots \wedge \hat{\textbf{w}}_{p}(t) \parallel,
\end{equation}
while here all the $k$ deviation vectors are normalized but we keep their directions intact. The general behaviour the GALI method (for different models and types of stability) as well as its predictability properties have been summarized recently in a review article \citep{SkoManLNP2016}. In short, and for TI systems the general evolution of the GALI$_k(t)$ is the following: (a) for chaotic orbits, it tends exponentially to zero with exponents that depend on the first $k$ LEs of the orbit while (b) for regular orbits it remains practically constant and positive (if $k$ is smaller or equal to the dimensionality of the torus on which the motion occurs) or it decreases to zero following a power-law decay (if $k$ is larger to the dimensionality of the torus on which the motion occurs; see e.g. \citealt{SkoBouAntPhyD2007,SkoBouAntEPJST2008}). Moreover, in \cite{ManSkoAnt} the behaviour and performance of  the GALI method was studied in the neighbourhood of invariant tori surrounding periodic solutions in the vicinity of periodic orbits in TD systems, where the role of sticky chaotic orbits and their diffusion properties were addressed in particular as well.

Aiming to capture and describe different dynamical time windows of the TD system, we calculate the GALI$_k$ as follows. Since our study refers to 3-dimensional configuration space (3-d.o.f. system), we will be using $k=3$ (i.e., 3 deviation vectors). The reason for choosing the $k$-value to be equal to the number of d.o.f. (as explained in several previous works; see e.g. \citealt{SkoBouAntPhyD2007,SkoManLNP2016}) is our goal to optimize the total computation time. This can be achieved by using the minimum number of deviation vectors (required to be calculated via the variational equations) which ensures the `safe' distinction between chaotic and regular motion at the same time. Then, by following the evolution of the index in time, one will record exponential decay of the GALI$_3$ for time intervals where the motion is chaotic while in any the other case it will refer to a non-chaotic one.

As explained more in more detail in \cite{ManBouSkoJPhA2013} and \cite{ManosMachado2014} the motion of an orbit in such a TD potential can be rather complicated and keep varying its dynamical stability from regular to chaotic and vice-versa when experiencing the changes of the phase space while the parameters of the potential change in time. So, if for example the deviation vectors of an orbit under study feel the chaotic dynamics of its regime for some time, then the volume formed by them will (see the definition of the GALI in equation~\ref{GALI:0}) shrink exponentially to very small values and remain small throughout the whole evolution unless one re-initializes the deviation vectors and hence their volume. Only then, they will be able to manifest the current new chaotic (again) or regular dynamics.

In general, we perform two slightly different procedures in the GALI$_3$ calculation, namely:
{\bf (i)} whenever we are interested in understanding more global dynamical trends, like for example how the amount of chaotic motion of a given galaxy component varies as a function of time in the TD system, we split the total time of integration in four fixed time intervals and we re-initialize the deviation vectors only in the beginning of each one. We employ this procedure for all the large samples of disc and halo orbits studied later on. Hence, we consider such time windows where the GALI$_3$ has enough time (with respect to the time scale of the total system evolution) to capture the chaotic or not motion of the orbit under study; and {\bf (ii)}  whenever we wish to plot a sample of orbits and show its detailed evolution in time, we let the GALI$_3$ evolve and whenever it reaches very small values (i.e.~GALI$_3 \leq 10^{-8}$) we re-initialize its computation by taking again $k=3$ new (always) random orthonormal deviation vectors, which resets the GALI$_3=1$. We then allow these vectors to evolve under the current dynamics. Let us stress that this procedure has been followed only in cases where we wanted to depict individual orbits and show some special characteristics of the motion accompanied by the chaos detection tool (GALI and/or MLE). These calculations of the GALI$_3$ were done separately, i.e. we have rerun smaller samples of typical orbits which illustrate several morphological and dynamical properties associated with the general trends shown together.

\subsection{Testing the equilibrium of the initial conditions} \label{test}

In the following sections, we will study the behaviour of ensembles of orbits and in particular we will measure their percentages of chaotic motion. This will be done by taking the coordinates of a sample of initial conditions from the beginning of the $N$-body simulation, and then evolving them in the presence of the time-dependent analytical potential.

The analytical potential is a good but imperfect representation of the actual potential experienced by the $N$-body particles. If the samples of initial conditions are not exactly in equilibrium with the analytical potential, this raises the concern that spurious chaotic motion could be introduced merely due to this presumed mismatch, particularly in the early stages of the evolution.

To investigate this possibility, we have performed the following test. We selected a random sub-sample of 1000 disc initial conditions from the beginning of the simulation. We then evolved them under three distinct situations for 2.5 Gyr: (1) The initial conditions are evolved directly into the time-dependent analytical potential; (2) The initial conditions are evolved in a frozen potential, i.e. in the time-independent analytical potential fixed at $t=0$; (3) The output of case 2 is now used as new initial conditions in the presence of the time-dependent analytical potential.

The evolution in the frozen potential of case 2 conserved energy to within $\Delta E / E \sim 9 \times 10^{-9}$. One would expect this procedure to dissipate eventual transients caused by lack of equilibrium. However, we measured the percentages of chaotic motion in the three cases and found that they are essentially the same, differing by less than 2.8 percentage points. For the purposes of our analyses, which aim to investigate global dynamical trends, this margin is certainly tolerable.

The similarity of results between cases 1 and 2 reinforces the notion that there is no serious departure from equilibrium that might compromise the results. Furthermore, the agreement between cases 1 and 3 indicates that the initial conditions can be run directly into the time-dependent analytical model, without prior application of step 2. By running such tests, we were unable to find indications that the percentages of chaotic motion could be severely overestimated. Moreover, the initial conditions seem to be in equilibrium within the potential to a fairly decent degree. For these reasons, we adopt method 1 directly in the remainder of the paper.

In our approach, the potential varies drastically during this first 2.5 Gyr where one of the main model components, i.e. the bar, starts to appear. This causes quite radical changes in the phase space which in turn have a strong impact in the dynamical stability of the periodic orbits surrounded by islands which may grow or shrink and/or (dis)appear simultaneously in the course of time. Moreover, how our initial conditions are distributed in the phase space plays a significant role as well as their corresponding energy \citep[see][for more discussion]{ManosMachado2014}. Hence, the most significant reason for observing such a large fraction of chaotic motion in the first epoch is due to these rather strong dynamical effects taking place while the bar is forming.

%%%%%%%%%%%%%%%%%%%%%%%%%%%%%%%%%%%%%%%%%%%%%%%%%%%%%
\section{The stellar disc} \label{sec:disc}

\subsection{Fractions of chaotic motion}

%-----------------------------------------------------
\begin{figure}
\begin{center}
\includegraphics[width=\columnwidth]{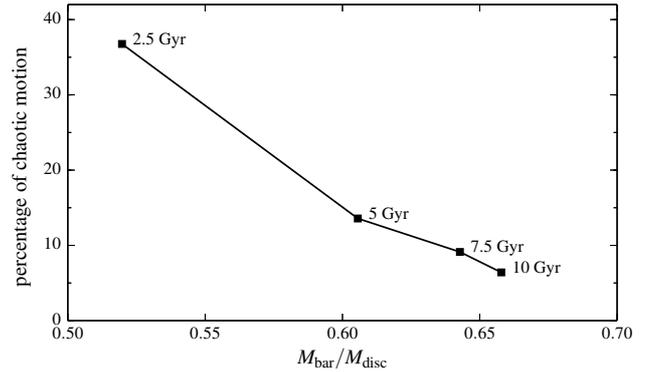}
\caption{Fraction of chaotic motion within the bar as a function of relative bar mass.}
\label{fig01}
\end{center}
\end{figure}
%-----------------------------------------------------

%-----------------------------------------------------
\begin{figure*}
\includegraphics[width=\textwidth]{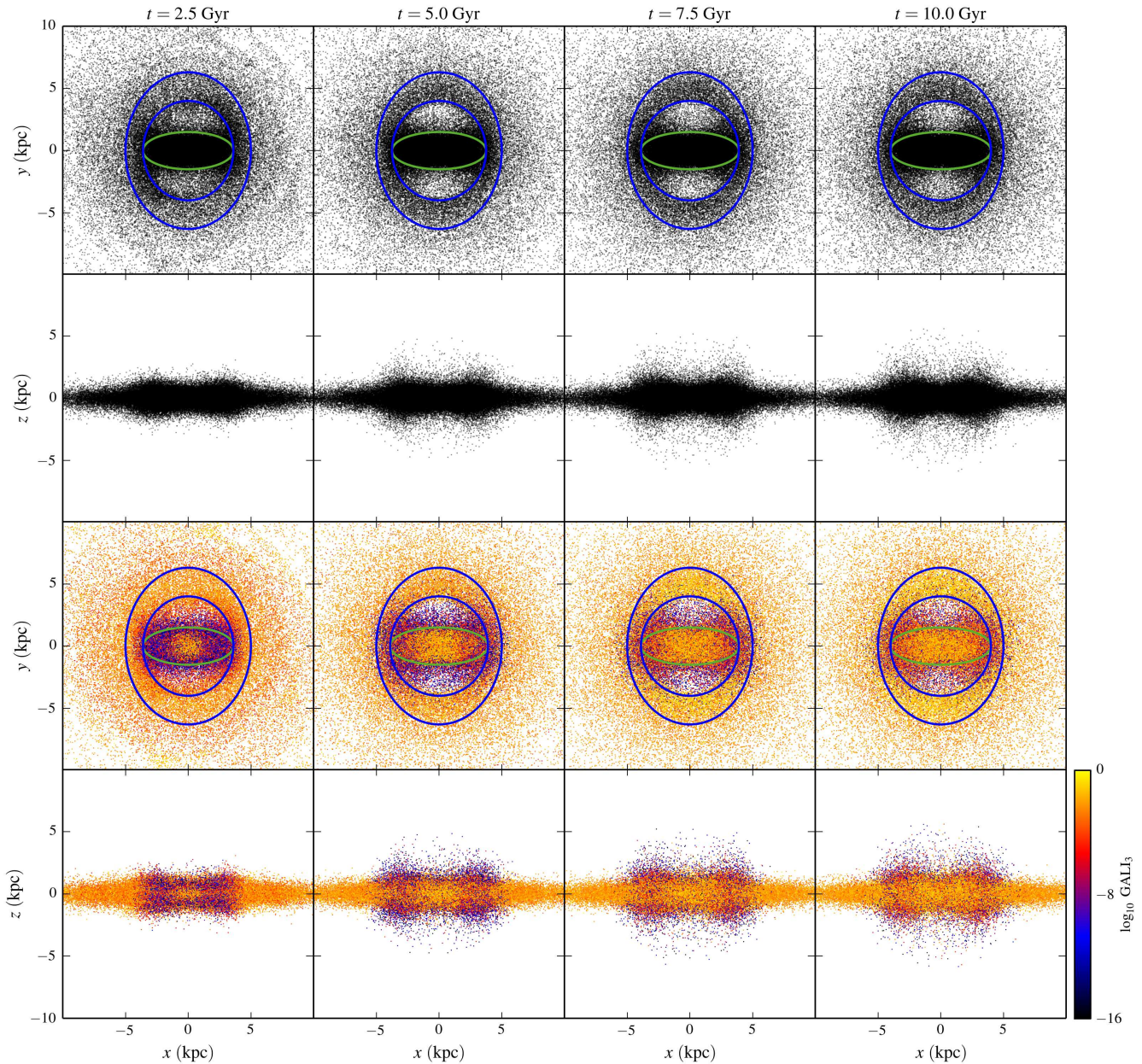}\\
\caption{Upper rows: face-on and edge-on views of the ensemble of orbits at the end of each time window. Lower rows: face-on and edge-on views of projected GALI$_{3}$ indices for the disc ensemble. Chaotic orbits are those with GALI$_{3} \leq 10^{-8}$. The face-on views also display the ellipses used to define the regions referred to as: bar, ring, gap and outer disc. Each frame is 20 kpc wide, and the particles are displayed in the reference frame that rotates with the bar.}
\label{fig02}
\end{figure*}
%-----------------------------------------------------

% correlation
In \cite{ManosMachado2014} we had seen that the overall fraction of chaotic motion in the disc decreases as the bar grows stronger. Figure~\ref{fig01} indicates the tight correlation between the decrease in chaos and the growth of the bar, as indicated by the fraction of chaotic motion -- measured within the bar -- as a function of relative bar mass (i.e. the bar-to-disc mass ratio). One also notices that most of the disappearance of chaos takes place during the first half of the evolution, which is the period of more intense bar growth.

% define regions
If bar growth is partially driving the rise of regular motion, the question then arises as to the spatial distribution of regular versus chaotic motions. In which regions of the stellar disc are regular and chaotic motion predominantly found? Is the global rise of regularity accompanied by some spatially localized increase of chaos? In order to address such questions, we define four distinct morphological regions: (i) the bar; (ii) the ring; (iii) the intermediate low-density region between the bar and the ring, referred to as the \textit{gap} region, for brevity; and (iv) the outer disc. These regions are selected somewhat arbitrarily (via visual inspection of the morphology), but they do reflect distinctive structural components, regarding density, as can be seen in the two upper rows of Fig.~\ref{fig02} with face-on and edge-on views, where the ellipses used to define them are shown.\footnote{In Fig.~\ref{fig02} and in all other such projections, the particles and orbits are displayed in the reference frame that rotates with the bar. Thus, the bar major axis always lies along the direction of the $x$-axis.}

% ensemble and gali3
To explore the spatially resolved evolution of chaos throughout the stellar disc, we resort to the analysis of an ensemble of orbits. From the $N$-body simulation, we select a sample of $1 \times 10^5$ disc particles at the time $t_0=1.4$~Gyr where the bar has already started to be formed and starts growing from that point on. Then, their coordinates are used as an ensemble of initial conditions to be evolved in the presence of the time-dependent analytical potential. We evolve these orbits for $10$~Gyr and study their dynamical behaviour. In order to avoid confusion, from now on we reset the $t_0$ to be zero (starting point of our simulations). Following \cite{ManosMachado2014}, we divide the total integration time in four intervals of $\Delta t = 2.5$ Gyr, re-initializing the GALI$_3$ index at the beginning of each window. The orbit is considered regular (non-chaotic) if its GALI$_3$ remains greater than $10^{-8}$ during a given time window; and it is considered chaotic if it reaches GALI$_3 \leq 10^{-8}$. In this manner, we are able to compute fractions of chaotic motion within each time window. Additionally, at a given instant in time, we can also compute spatially resolved chaos fractions in different regions of the disc.

% yellow-purple maps
A global picture of the spatial distribution of regular and chaotic motion in the disc can be seen in the two bottom rows of Fig.~\ref{fig02}, which displays the face-on and edge-on views of the ensemble of disc particles at the end of each time window, coloured by the GALI$_{3}$ index (being chaotic towards the blue, and regular towards the yellow). Some major results are already noticeable even by eye. First, the striking decrease of chaos within the bar region can be clearly seen. Secondly, even though the gap is a very low density region, it seems to hold a good portion of the chaotic orbits. Third, the outer disc -- as well as the ring, to a degree -- seem quite dominated by regular motion. Finally, another outstanding feature is the peanut-shaped view of the bar seen in the edge-on projection (sometimes called X-shaped bulge). Remarkably, particles that depart considerably from the $z=0$ plane are mostly chaotic.

\begin{figure}
\begin{center}
\includegraphics[width=\columnwidth]{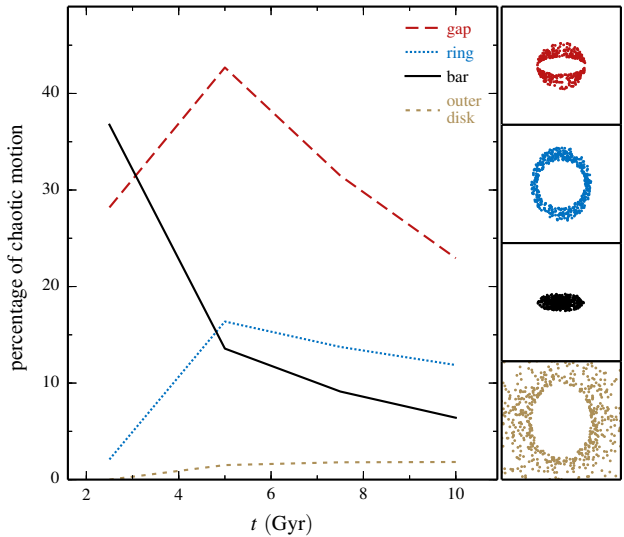}\\
\caption{Fraction of chaotic motion as a function of time, measured within different regions of the disc. These regions are schematically indicated in the right-hand panels, and more clearly detailed in Fig.~\ref{fig02}.}
\label{fig03}
\end{center}
\end{figure}

% quantify fchaos in regions
In order to quantify in more detail these results, we measure the fraction of chaotic orbits as a function of time in each region (i.e. at the end of each time window, we obtain the number of particles having GALI$_{3} \leq 10^{-8}$ in a region divided by the total number of particles within that region). The result is shown in Fig.~\ref{fig03}. The fraction of chaotic motion within the bar drops from nearly 40 per cent to less than 10 per cent. The outer disc remains essentially regular, with a non-zero but negligible appearance of chaos throughout the evolution. The fact that the ring region undergoes an initial increase in chaos can be ascribed in good measure to the edges of the bar. The gap region displays some interesting behaviour. Between the first and second time windows, the gap becomes depleted in terms of total number of particles, but at the same time its fraction of chaotic motion increases. From then on, it decreases, but the gap continues to be the region holding the highest local fraction of chaotic motion in the disc. The large amount of chaotic motion seen in the gap region is not unexpected. In fact it is well known that orbits that oscillate between the Lagrangian points $L_1$ and $L_2$ are unstable and therefore the transition zone between the bar and the disk is expected to be chaotic \citep[e.g.][]{HarKalMNRAS2009, AthaRomMas2009MNRAS}.

%-----------------------------------------------------
%-----------------------------------------------------

\subsection{Morphology and evolution}

% changes of state
In the analyses of the previous section we considered the state (chaotic or regular) of each particle at specific instants -- the ends of the four time windows. Now we will consider the changes of state. For example, one given orbit that was reckoned to be regular at the end of the evolution might have been chaotic at the beginning, or it might have been continuously regular. In either case, where did such particles originate? Do the particles that change dynamical state (and those that don't) share a common locus at the beginning of the evolution? To explore these issues, we will examine separately the orbits that change dynamical behaviour and those that do not. This will allow us, in a sense, to map the origins and the destiny of regular and chaotic motion.

\begin{figure*}
\begin{center}
\includegraphics[width=\textwidth]{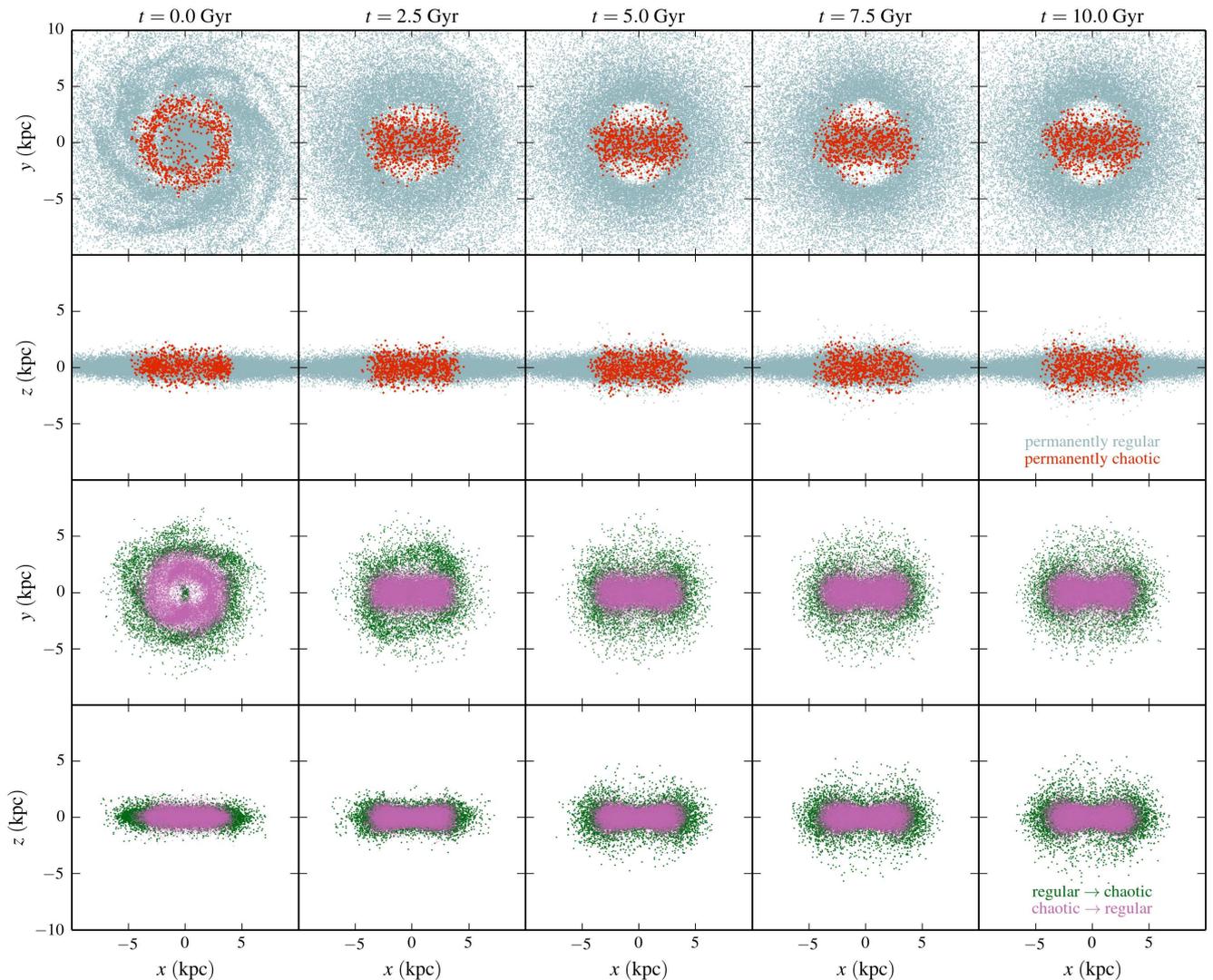}\\
\caption{Evolution of regular/chaotic regimes separated in different configurations. Upper rows: face-on and edge-on vies of the orbits that are permanently regular (cyan), and those that are permanently chaotic (red). Bottom rows: face-on and edge-on views of the orbits that start regular and end chaotic (green), and those that start chaotic and end regular (purple).}
\label{fig04}
\end{center}
\end{figure*}

% permanent
Let us start by selecting those orbits which are permanently regular (64.3 per cent) and those which are permanently chaotic (only 0.9 per cent). The remainder (34.8 per cent) change their nature at least once during the evolution. Let us consider first those orbits that do not undergo any change of regime throughout the evolution (upper rows of Fig.~\ref{fig04}).
%-----------------------------------------------------
%-----------------------------------------------------

\textit{Orbits that are permanently regular.} As regards morphology, the permanently regular ones are qualitatively unremarkable, in the sense that they occupy almost any region of the galactic disc. There is thus little qualitative distinction between them and the entire ensemble and they merely map the normal evolution of the galactic disc as a whole. The only noticeable structures that are not quite covered by these orbits are the gap region, and, vertically, the peanut. (Due to the method employed to create the initial conditions in the $N$-body simulation of  \citealt{MachadoAthanassoula2010}, there is a residual transient seen as a vague spiral pattern at $t=0$ and it subsides on a short time scale.)

\textit{Orbits that are permanently chaotic.} The permanently chaotic orbits, on the the other hand, display peculiar features. They are tightly restricted to the region of the bar, and partially to the gap. Indeed, they spend nearly the entire evolution confined within this region. There is not one single permanently chaotic orbit to be found in the outer regions of the disc. At the instant $t=0$, these particles -- whose future destiny is to be permanently chaotic -- are initially located within a reasonably well-defined ring, i.e. they are mostly found within $2\,{\rm kpc} < r < 4\,{\rm kpc}$.

% changing state
For the remainder of the orbits, we will focus on two regimes: those that start regular and end chaotic (6.1 per cent), and those that start chaotic and end regular (20.8 per cent), regardless of the intermediate states (i.e. the transitions in the second and third time windows). Finally, there is a subset of orbits (7.9 percent) that do undergo two changes of regime, but nevertheless finish as they started; these will be disregarded. Let us consider now the two cases where the final state differs from the initial state (lower rows of Fig.~\ref{fig04}):

\textit{Orbits that begin regular and end chaotic.} These start at $t=0$ from a similar locus as the permanently chaotic, but here the ring is slightly larger and more diffuse. This subset also includes some orbits very close to the origin ($r<0.5$ kpc) in the beginning, which are not present in the permanently chaotic case. The volume occupied by these orbits contracts gradually, but they are more extended than the permanently chaotic ones, encompassing the region of the gap at later times as well. They are also vertically extended, being the major contributors to the structure of the peanut. In fact, this is the only subset of particles which significantly populates the peanut, in the regions of about $2.5\,{\rm kpc} < |z| < 5\,{\rm kpc}$ of height. The gap region, and mainly the peanut, are regions where chaos is important. However, it is only the initially regular -- and finally chaotic -- orbits that depart considerably from the plane. The orbits that were already chaotic from the beginning do not visit such heights.

\textit{Orbits that begin chaotic and end regular.} Interestingly, the initial locus of this subset is approximately the complement of the previous case. Here, the orbits at $t=0$ occupy the region internal to the ring defined by the previous case, while avoiding the very centre. In the third row of Fig.~\ref{fig04}, the purple points overlap green points in the $t=2.5-10$ Gyr frames. But in the $t=0$ frame, the purple points fill precisely an empty region. Subsequently, the initially chaotic orbits evolve to be essentially part of the bar and end regular.

%-----------------------------------------------------
\begin{figure*}
\includegraphics[width=\textwidth]{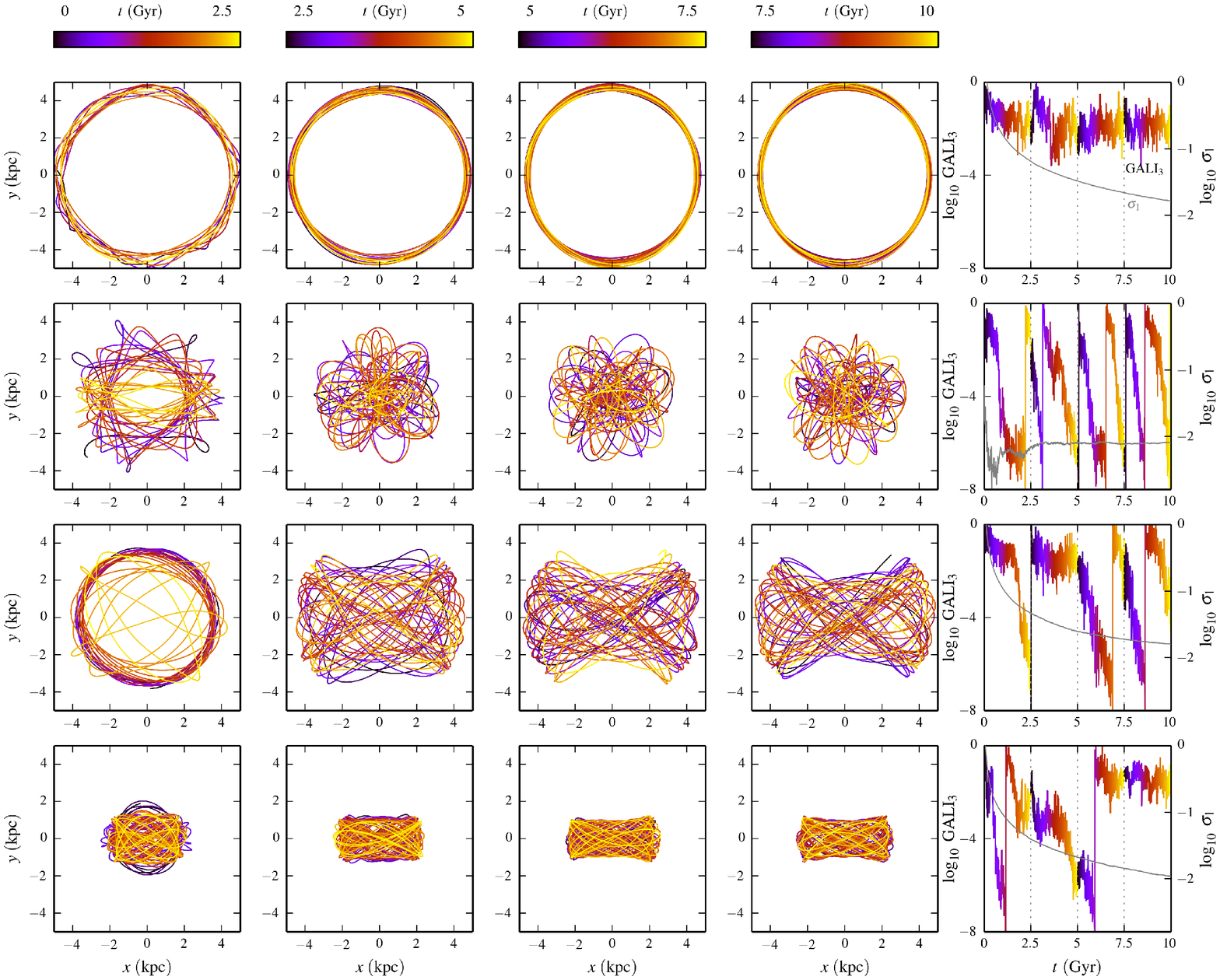}\\
\caption{Typical examples of disc orbits. Each row displays one orbit at four time windows, coloured by time (from black to yellow at each window). The fifth column shows the corresponding GALI$_{3}$ evolution, as well as the MLE $\sigma_1$. First row: a permanently regular orbit. Second row: a permanently chaotic orbit. Third row: an orbit that begins regular and ends chaotic. Fourth row: an orbit that begins chaotic and ends regular. The orbits are displayed in the reference frame that rotates with the bar.}
\label{fig05}
\end{figure*}
%-----------------------------------------------------

In Fig.~\ref{fig05}, we show four representative examples of disc orbits that illustrate the dynamical transitions described above. Note that each row displays one given orbit whose time evolution has been divided in four parts. In more detail, these orbits are a typical sample of initial conditions that correspond to the four regimes depicted in Fig.~\ref{fig04}. Each row here displays one orbit at four time windows, coloured by time (from black to yellow at each window). The fifth column shows the corresponding GALI$_{3}$ evolution (note that here GALI$_{3}$ are re-initialized whenever they reach the value $10^{-8}$). Also shown in the fifth column are the MLE $\sigma_{1}$.
In the first row of Fig.~\ref{fig05}, one sees a permanently regular disc orbit (corresponding to one of the cyan particles of Fig.~\ref{fig04}). This orbit does not change its regular dynamical nature, nor its morphology throughout the total time evolution. Its GALI$_3$ fluctuates around a positive value until the end of the integration.
In the second row of Fig.~\ref{fig05}, we show an example of a permanently chaotic orbit (corresponding to one of the red particles of Fig.~\ref{fig04}). Its  GALI$_3$ decays exponentially to zero consecutive times, and its morphology is distinctively irregular throughout.
In the third row of Fig.~\ref{fig05}, we plot an orbit that begins regular and ends chaotic (corresponding to one of the green particles of Fig.~\ref{fig04}). This orbit initially has a roughly circular morphology but subsequently loses its regularity and becomes more elongated along the direction of the bar. Note its GALI$_3$ evolution: in the very first part of the evolution the index remains non-zero before starting to decay exponentially to zero. This first drop is accompanied by the onset of irregularity, clearly seen in the first panel of this orbit, as the nearly circular motion (black at the beginning of the time window) already starts giving way to an irregular morphology (yellow at the end of the time window). Let us here recall that this orbit was initially characterized as regular (non-chaotic to be more accurate) when employing a global run where we were registering the final value of the  GALI$_3$ at the end of each time window, and from this we derive the conclusion that at least for that time interval the orbit exhibits more regular behaviour than chaotic (as its projections supports as well). However, in this figure we always reset the  GALI$_3$ only when it strictly crosses the threshold, as mentioned before.
Finally, in the fourth row of Fig.~\ref{fig05}, we display an orbit that begins chaotic and ends regular (corresponding to one of the purple particles of Fig.~\ref{fig04}). This orbits begins strongly chaotic, as indicated by its GALI$_3$, which drops exponentially to zero after only a rather small number of  iterations. Furthermore, its stability is changing drastically in time and it is becoming gradually regular; its GALI$_3$ at the last part of the evolution fluctuates around a positive value. Moreover, from a morphological point of view, it is being transformed into a bar orbit under the effect of the TD potential.
% MLE discussion
In the fourth column of Fig.~\ref{fig05}, besides the GALI$_3$, we also display the MLE in lin-log scale (grey). For the two upper panels, things are quite clear: for the permanently regular orbit (first row) the MLE decays to zero following a power law (which is even more clear in a log-log scale, not shown here) while for the permanently chaotic (second row) it tends to a positive value. However, whenever an orbit experiences more complex dynamical transitions, like the orbits shown in the third and fourth rows, the MLE, as an averaging measure, faces several difficulties, described also in \cite{ManBouSkoJPhA2013,ManosMachado2014}. Starting with the fourth row (an orbit that begins chaotic and ends regular), we may see that the MLE does not have sufficient time to capture the first chaotic period of the orbit. In the third row (an orbit that begins regular and ends chaotic), at first glance the MLE does not seem to describe accurately the last half or so of the orbital evolution, since in the lin-log scale it is not clear that there is a tendency to approach a positive value. For this reason we have checked its evolution in log-log scale (not shown here), where a change in the slope becomes clearer as it approaches a non-zero value, indicating the dynamical change of the orbit. However, it is already evident that one should integrate for much longer periods to get a convincing information from the MLE.

Let us here discuss into more detail these orbits that start regular and end chaotic (green points in Fig.~\ref{fig04}). These orbits seem to populate eventually (and mostly) the gap region around the Lagrangian $L_1$ and $L_2$ points. Thus, it is a reasonable to wonder whether there is a relation between this transition to chaoticity with the stickiness that characterizes this region. As a general and typical trend, when observing the evolution of a sticky chaotic orbit, one sees a trajectory that has a regular-like morphology for some time before eventually revealing its chaotic nature (when entering a strongly chaotic regime). Typically the morphology of these orbits is preserved, in the sense that e.g. disc orbits remain disc orbits and the bar ones remain bar orbits, but surely with a rather different rate of diffusion in configuration and phase space in the two different epochs. In our case, a large fraction of `regular $\rightarrow$ chaotic' orbits (green points in Fig.~\ref{fig04}) are trajectories that additionally transform their spatial morphology (mostly from disc to bar orbits). Such a typical example is given in Fig.~\ref{fig05} (third row), where an initially almost spherical (regular) disc orbit gradually changes to a bar (chaotic) orbit. This transition is mostly due to the emergence of the bar component in the potential as the time evolves and it does not seem to be strongly related to stickiness effects as in other cases studied broadly in the literature. Hence, a lot of the transitions from chaotic to regular and/or vice-versa are due to the changes of the stability of certain regions in the phase space.

%-----------------------------------------------------
\begin{figure*}
\includegraphics[width=\textwidth]{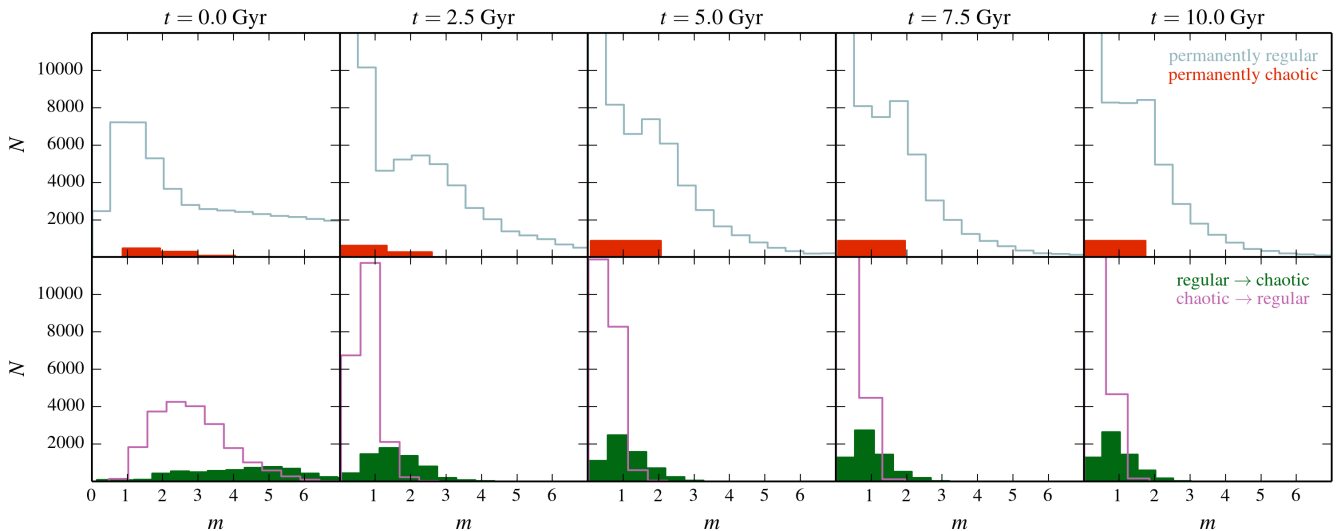}
\caption{Histograms of positions of disc particles in terms of the dimensionless quantity $ m^2 =  (x/a)^2 + (y/b)^2 + (z/c)^2 $, where $a,b$ and $c$ give the shape of the bar at each time. Particles within the bar have $m<1$. Top row: permanently regular (cyan) and permanently chaotic (red). Bottom row: particles that begin regular and end chaotic (green) and particles that begin chaotic and end regular (purple).}
\label{fig06}
\end{figure*}
%-----------------------------------------------------

In order to investigate further the morphological properties of these different dynamical transitions, we then examined the distribution of each group of orbits as a function of the positions\footnote{We have also inspected these distributions in terms of the spherical radius $r$ or cylindrical radius $R$ but they are less clear regarding the bar region.} in terms of the dimensionless quantity $m^2 =  (x/a)^2 + (y/b)^2 + (z/c)^2$, meaning that particles within the bar have $m<1$. The three shape parameters $a,b$ and $c$ are time-dependent and are looked up at each desired instant to compute $m$. In the top row of Fig.~\ref{fig06}, we show the histograms of such positions of disc particles for the permanently regular (cyan) and permanently chaotic (red). In the bottom row, we plot the corresponding histograms for the particles that begin regular and end chaotic (green) and particles that begin chaotic and end regular (purple).

Starting with the top row, we notice that the absolute number of permanently chaotic orbits is small, and that they are located essentially within the bar or near its edges, at least after $t=2.5$ Gyr. Surely, in the same ranges one may find far more regular motion as well; yet they are more widely spread in the several distances compared to the chaotic ones. The most striking feature of Fig.~\ref{fig06} is the fact that the purple histograms (chaotic $\rightarrow$ regular) are almost completely confined to the bar region ($m<1$) at all times after  $t=2.5$ Gyr. The green histograms (regular $\rightarrow$ chaotic) show a tendency in the same direction, but less pronounced, and there are far fewer such orbits in absolute numbers. Complementing the information provided by Fig.~\ref{fig03} and Fig.~\ref{fig04}, these histograms offer a more quantitative overview regarding the spatial distribution of these types of motion, highlighting not only the preferred locations where each type of regime is to be found, but also their relative contributions in number of particles. For example, the comparisons of the green and purple histograms confirm the visual impression of Fig.~\ref{fig04} -- that the finally regular orbits become more confined and the finally chaotic becomes more spread out -- but also displays quantitatively that the finally regular are far more prevalent. Also, we see that the initially chaotic and the permanently chaotic types both spawn from roughly the same region, although the latter is more rare.

%%%%%%%%%%%%%%%%%%%%%%%%%%%%%%%%%%%%%%%%%%%%%%%%%%%%%
\section{The dark matter halo} \label{sec:halo}

Having characterized the evolution of the fraction of chaotic motion in the disc, we now turn to the dark matter halo, to evaluate how halo orbits behave in the presence of an evolving barred galaxy.

% ensemble and gali3
Analogously to the disc analysis, we now select a sample of $2 \times 10^5$ halo particles from the $N$-body simulation. This ensemble is also evolved in the presence of the time-dependent analytical potential, and we use the GALI$_3$ index to study the fraction of chaotic motion in the halo (in a similar manner as we did for the disc particles), exploring both the time evolution and the spatial distribution.

\subsection{Inner halo} \label{inner}

%-----------------------------------------------------
\begin{figure}
\includegraphics[width=\columnwidth]{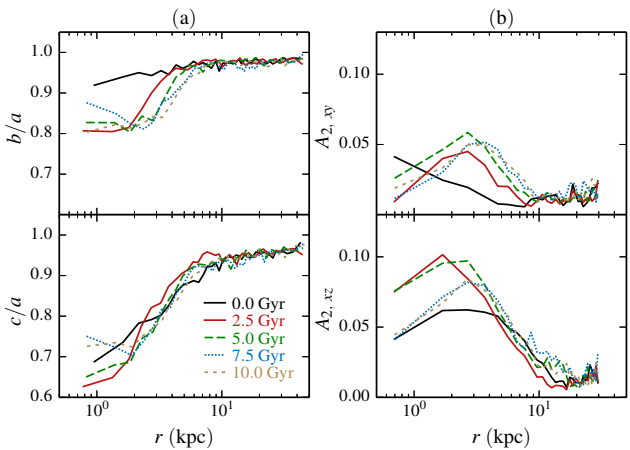}
\caption{Shapes of the ensemble of halo orbits at different times: (a) Axis ratios $b/a$ and $c/a$ as a function of distance from the centre; (b) Relative amplitude of the $m=2$ Fourier component of the mass distribution projected on the $xy$ and $xz$ planes.}
\label{fig07}
\end{figure}
%-----------------------------------------------------

%-----------------------------------------------------
\begin{figure*}
\includegraphics[width=\textwidth]{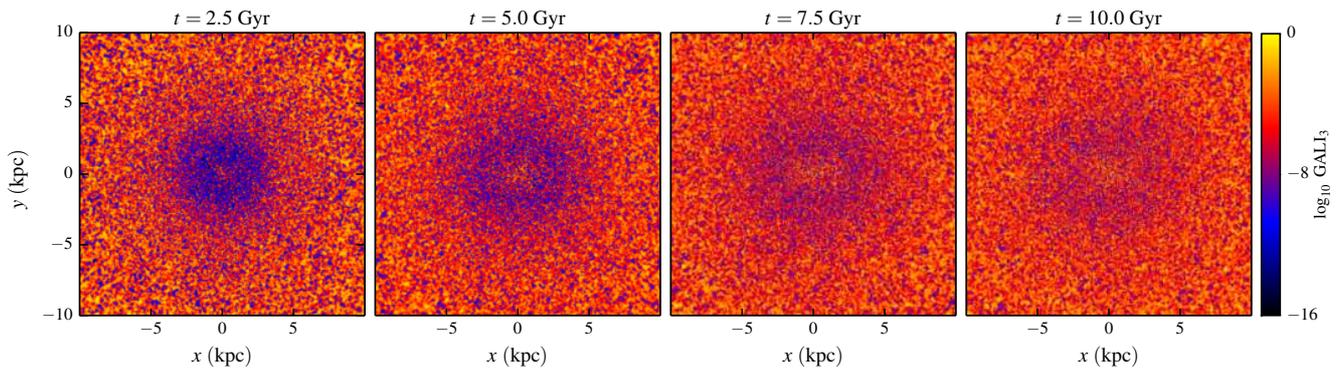}
\caption{Projected GALI$_{3}$ values (in logarithmic scale) for the halo ensemble, at the end of each time window. Chaotic orbits are those with GALI$_{3} < 10^{-8}$. Each frame is 20 kpc wide.}
\label{fig08}
\end{figure*}
%-----------------------------------------------------

%-----------------------------------------------------
\begin{figure}
\includegraphics[width=\columnwidth]{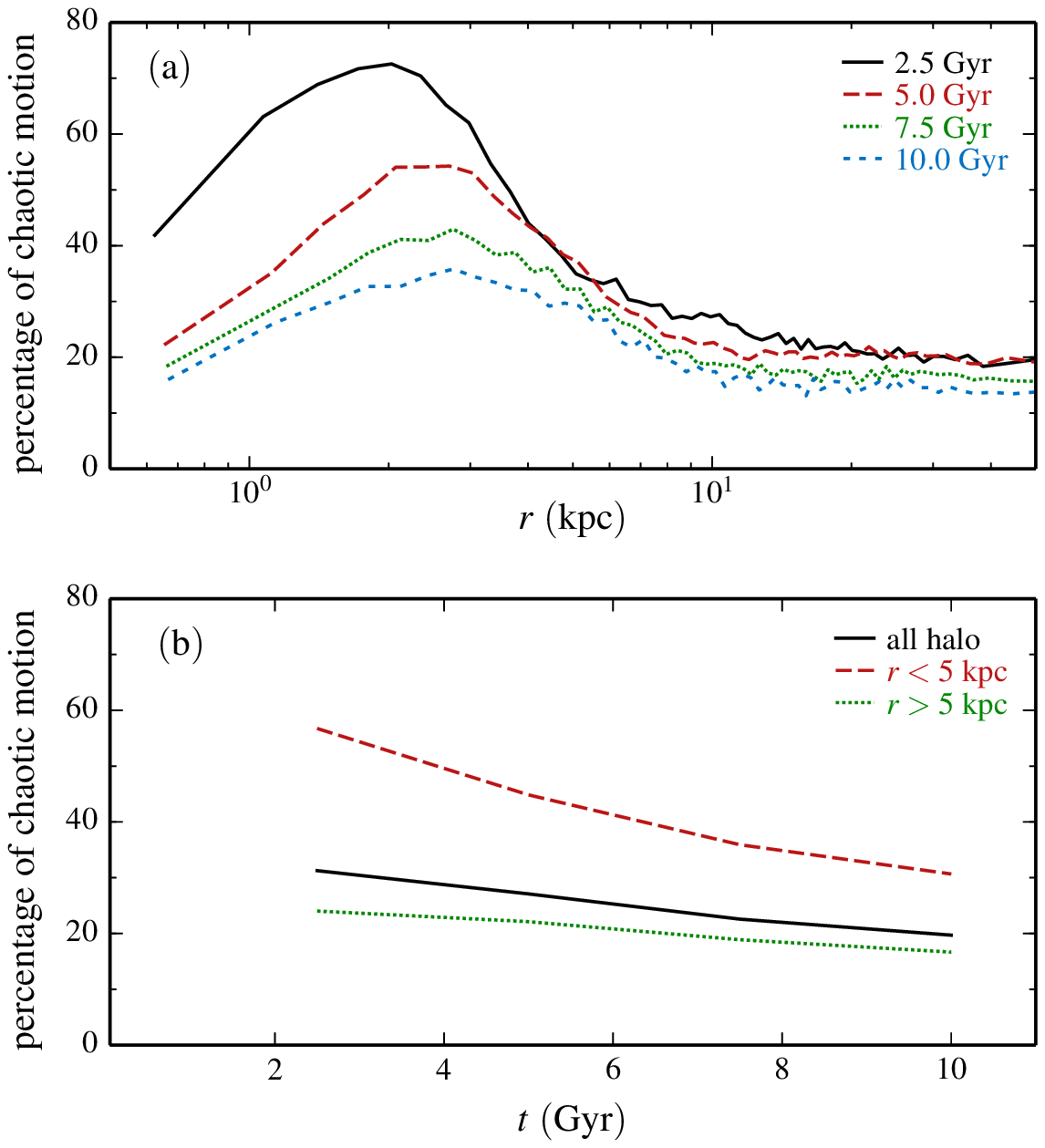}
\caption{Fraction of chaotic motion in the halo: (a) as a function of radius for different times, and (b) as a function of time for different regions.}
\label{fig09}
\end{figure}
%-----------------------------------------------------

% reproduce features
The ensemble of halo initial conditions behave as massless test particles and cannot be expected to mimic exactly the self-consistent gravitational evolution of the $N$-body simulation. Nevertheless, global morphological features are reproduced with quite good agreement. That was also the case for the ensemble of disc initial conditions, as shown in \cite{ManosMachado2014}, and it is one of the indications that our time-dependent analytical model provides a remarkably useful approximation of the galactic potential, for several purposes.

% halo bar
A distinctive halo feature which is reproduced is the so-called \textit{halo bar}, \citep{Athanassoula2005, Athanassoula2007, MachadoAthanassoula2010}, also called the \textit{dark matter bar} \citep{Colin2006}. In $N$-body simulations, the inner halo of strongly barred galaxies is found to become prolate. This halo bar is a structure that rotates together with the disc bar, but it is not as strong and not as elongated as its stellar counterpart. We found that this peculiar feature also arose from the ensemble of halo initial conditions in the presence of our analytical model. One should bear in mind that the halo analytical potential remains spherically symmetric by construction throughout the evolution. The bar potential is thus responsible for inducing the prolateness of the inner halo orbits.

% shapes
Here we will characterize the halo bar, using the ensemble of initial conditions and analysing their morphological evolution. The reader should notice that in the present context, when we speak of `the halo' and its shape, we mean the ensemble of orbits within the analytical potential. To quantify the shape of the inner halo, we measure the axis ratios $b/a$ (intermediate-to-major) and $c/a$ (minor-to-major) as a function of distance from the centre (along the major axis). This is done employing the same techniques as in \cite{MachadoAthanassoula2010}: we compute the axis ratios using the eigenvalues of the inertia tensor. To avoid the bias of spherical shells, the particles are sorted as a function of local density, and the shapes are measured inside density bins containing equal number of particles. In this manner, the spheroidal bins naturally follow the shapes of the isodensity surfaces. The resulting shape profiles are shown in Fig.~\ref{fig07}a. The outer shape or the halo (certainly beyond $r \gtrsim 10$ kpc) is globally spherical, as one would expect; both axis ratios reach values of 0.95--1, i.e. as spherical as possible within the resolution set by the particle number. In the inner region, however, $b/a$ drops to $\sim$0.8--0.85 and $c/a$ to $\sim$0.6--0.75. This means a halo bar which is non-circular on the plane of the disc, and even more flattened in the vertical direction. The relatively steep drops take place at around $\sim$5 kpc, which could be regarded as the approximate size of the halo bar.

% fourier
Complementary, another way of quantifying the shape of the inner halo is to use the Fourier components of the bidimensional mass distribution, namely the relative amplitude of the $m=2$ component. This is often used to quantify bar strengths, as it is quite sensitive to departures from axisymmetry. In Fig.~\ref{fig07}b, we show this relative $A_2$ amplitude measured by projecting the particles on the $(x,y)$ plane, or else on the $(x,z)$ plane. The amplitudes are surely much less pronounced than those of stellar bars, but the elongated and flattened morphology of the halo bar is clearly measurable, indicating its bar-like nature. Here we see that $\sim$5 kpc marks roughly mid-height between the $A_2$ peak and the first minimum, so to speak.

% why
The purpose of characterizing the halo bar is twofold. First, it helps underscore the efficacy of the time-dependent analytical model developed in \cite{ManosMachado2014}, which is able to recover even such a detailed morphological feature that was not deliberately built into the model. Secondly, it provides a well motivated distinction between `inner' and `outer' halo in the present context. Contrary to the stellar disc, the halo has no obvious morphological substructures, but the existence of the halo bar suggests a natural separation. We proceed to study the fractions of chaotic motion in the halo as a whole, and within the inner and outer parts.

%-----------------------------------------------------
\begin{figure*}
\includegraphics[width=\textwidth]{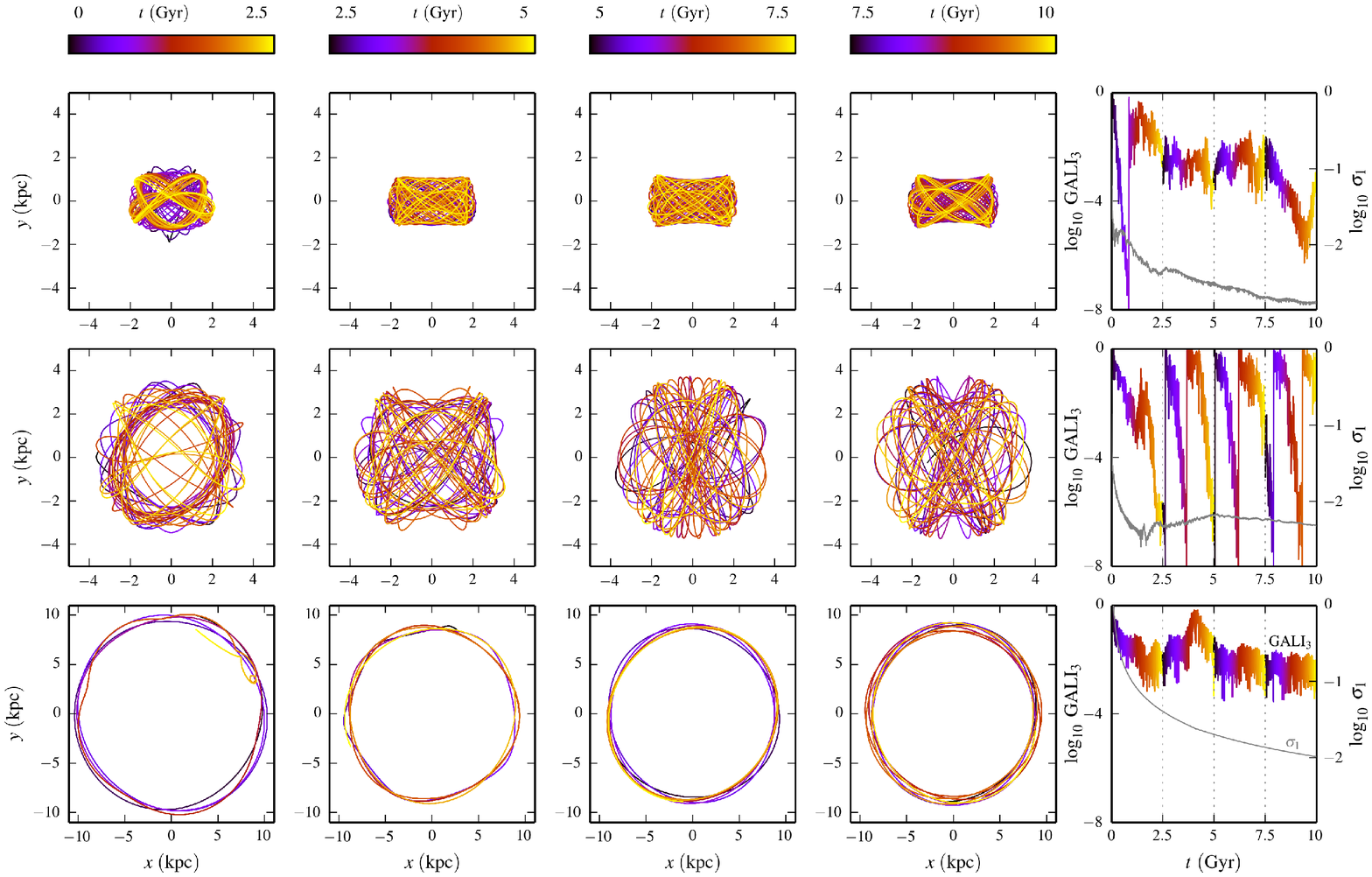}
\caption{Examples of halo orbits. Each row displays one orbit at four time windows, coloured by time (from black to yellow at each window). The fifth column shows the corresponding GALI$_{3}$ evolution, as well as the MLE $\sigma_1$. Notice the different spatial scales: the top and middle rows are orbits from the innermost parts of the halo, while the bottom row is an orbit from a region ($r \sim 10$ kpc) where the halo is essentially spherical. The orbits are displayed in the reference frame that rotates with the bar.}
\label{fig10}
\end{figure*}
%-----------------------------------------------------

\subsection{Fractions of chaotic motion in the halo}

% maps
In Fig.~\ref{fig08}, we show four snapshot maps where the colour at each point indicates the value of the GALI$_{3}$ (at a given momentary position) for the halo ensemble, at the end of each time window. Chaotic orbits are again those with GALI$_{3} < 10^{-8}$. Each frame here is 20 kpc wide. The measurement of the GALI$_{3}$ indices at the end of each time window (namely at $t =$ 2.5, 5, 7.5 and 10~Gyr) allows us to obtain the fraction of chaotic motion as a function of time. Furthermore, we are able to measure this fraction with spatial resolution. Figure~\ref{fig08} thus displays the prevalence of chaotic or regular orbits in the halo at four instants: here chaos is represented by the blue (darker) colors. From this figure we can already discern the two main results: (i) chaos decreases globally in the halo and (ii) chaos is only ever dominant in the inner region ($r < 5$ kpc).

% fractions
Now let us quantify these halo trends in more detail. For brevity we refer here to the expressions `inner' and `outer' halo in the sense defined in Section \ref{inner}, i.e. `inner' meaning essentially the region where the halo bar is found, and `outer' being all the remainder of the halo, which is nearly spherical. Figure~\ref{fig09}a displays the radial profile of the fraction of chaotic motion at the end of each time window. We see that at any given time, considerably more chaos is to be found in the inner halo than in the outer halo. Within $r < 5$ kpc, chaotic orbits even dominate at first. The time evolution of this fraction is shown in Fig.~\ref{fig09}b. In the outer halo, the amount of chaotic motion is always low ($\sim$ 20 per cent) and decreases slightly with time. If the halo is considered as a whole, the global fraction of chaotic motion decreases from 30 to 20 per cent, approximately. But the most pronounced evolution is seen in the region of the halo bar: from nearly 60 to 30 per cent. So the overall result is that the global fraction of chaotic motion in the halo decreases monotonically with time, and at any given radial range. Another result to be underscored is that in the inner halo, chaotic motion initially dominates over regular motion, i.e. more than 50 per cent, which was never the case in any of the disc regions.

The fact that chaotic orbits dominate in the inner halo is consistent with the findings of \cite{Valluri2013}, using a halo extracted from a cosmological hydrodynamical simulation. It would be interesting to understand whether the large percentages of chaotic orbits are due to scattering by the time-dependent bar potential (e.g. see \citealt{Price-Whelan2016} on the effect of a time-dependent bar on the Orphiuchus tidal stream) or if it is due to the deeper central potential. First, to verify that our method does not overestimate the initial amount of chaos, we performed the same test described in Section~\ref{test} also to a random sub-sample of 1000 halo initial conditions. The evolution of this sub-sample in a frozen potential conserved energy to within $\Delta E / E \sim 4 \times 10^{-5}$. Comparing the percentages of chaotic motion in this case, we found that in the frozen potential there was marginally less chaos (by 3.2 percentage points). That might lead us to speculate that the greater amount of chaotic motion is indeed due to the time-dependence of the bar potential. Furthermore, the first time window may witness more chaotic motion because that is the period during which all the potential parameters undergo their most intense variations. But to disentangle the role of the central concentration, a systematic comparison of models would be needed, which is beyond the scope of this paper.

\subsection{Sample of halo orbits}

% sample
From the ensemble used to study the halo, we select three orbits to illustrate characteristic behaviours. In Fig.~\ref{fig10}, each row corresponds to one orbit, and each orbit is shown at four time windows, coloured by time (from black towards yellow at each window). The fifth column exhibits the GALI$_{3}$ and the MLE evolution of each orbit. Note that here, the GALI$_{3}$ are reset once they reach $10^{-8}$, being then considered chaotic. The first row of Fig.~\ref{fig10} shows an example of an orbit that was chaotic at first, but then becomes regular later. This is seen both by the orbital shape itself and by the GALI$_{3}$ drop in the first time window. This orbit becomes part of the halo bar, as evidenced by its orientation, size and shape. In the fourth time window there is a mild backslide away from regularity, again discernible both in the morphology and in the GALI$_{3}$, but this momentary relapse is insufficient to reach the threshold of being classified as chaotic. In the second row of Fig.~\ref{fig10}, we present an example of an orbit that is permanently chaotic, since its morphology is consistently irregular and its GALI$_{3}$ drops repeatedly. Then, a typical orbit that is permanently regular, being even roughly circular and its GALI$_{3}$ never drops significantly, can be found in the third row of Fig.~\ref{fig10}. Notice that the first and second rows are orbits from the inner part of the halo, while the third example is an orbit from an outer region ($r \sim 10$ kpc), where the halo is essentially spherical.

%MLE discussion
One may notice that the dynamical transitions observed in this sample of halo orbits, shown in Fig.~\ref{fig10}, are generally simpler compared to those disc orbits shown in Fig.~\ref{fig05}, i.e. they maintain their dynamical stability for most of the time integration. That makes it easier for the MLE to describe accurately the chaotic or regular nature of this orbits in the TD potential. Thus, here we can see a good match between the GALI$_{3}$ and MLE.

%%%%%%%%%%%%%%%%%%%%%%%%%%%%%%%%%%%%%%%%%%%%%%%%%%%%%
\section{Summary and conclusions} \label{sec:conclusion}

We employed the time-dependent analytical model developed in \cite{ManosMachado2014} to perform a detailed analysis of the evolution of chaotic and regular orbits in a barred galaxy. The galaxy model is represented by a gravitational potential composed of three components -- disc, bar and halo -- whose parameters are all fully time-dependent and were derived from a self-consistent $N$-body simulation. By directly following the dynamical evolution of ensembles of orbits within the analytical potential, we were able to calculate the fractions of chaotic and regular motion resolved in both time and space. With this information we could evaluate not only the global trends in time, but also across several regions of the galactic disc and of the halo, associating them with distinct morphological features. We scrutinized the different changes of regime during the evolution, tracing the types of orbits back to their common origins.

Our previous study of the disc was now extended by associating simultaneously the dynamical and the morphological state of orbits. The time-dependence of the analytical model ensures rather realistic dynamical transitions similar to an $N$-body simulation, i.e. bar formation and growth, development of a ring, a dynamical halo, etc. At the same time, this setup serves ideally to apply the GALI chaos detection method and in this way to determine the current chaotic (or otherwise) dynamical state of any given orbit at a fixed time interval, something that it is extremely hard to do in an $N$-body simulation. Moreover, it has the advantage over a derived frozen potential from an $N$-body simulation, from the point of view that it incorporates smooth dynamical evolution via its time-dependent parameters and allows us to follow the both dynamical and morphological transitions.

We analysed the fractions of chaotic motion within different morphological components of the disc, namely the bar, the ring, the gap region between them, and the outer disc. Then we also investigated the origins of four different regimes of transitions between dynamical stability: those orbits that are permanently regular, those that are permanently chaotic, those that start regular and end chaotic, and those that start chaotic and end regular. Here we summarize these results pertaining to the disc:

\begin{enumerate}

\item We found that the overall trend in time is a global decrease of chaotic motion in the disc as a whole. In the bar region, the decrease is the most intense and it is monotonic, going from nearly 40 to less than 10 per cent of chaotic motion. The outer disk is overwhelmingly dominated by regular motion throughout the entire evolution, and its fraction of chaotic motion never exceeds 2 per cent.

\item The ring and the gap show more complicated evolutions. The gap region is most peculiar in the sense that it is a very low density region, but although it is sparsely populated, its fraction of chaotic motion is the highest seen in the disc, reaching more than 40 per cent at one point. The ring also differs from the rest of the disc regions because it undergoes a net increase of chaotic motion, locally.

\item Regarding the orbits that do not change their dynamical state during the evolution, we saw that permanently regular orbits may be found nearly anywhere in the disc. The permanently chaotic orbits, however, are only found within the bar, and their origin can be traced back to a confined annular region at $t=0$, whose average radius is comparable to the future length of the bar.

\item Regarding the orbits that do change their dynamical state between the first and the last time windows, we see that the ones that start chaotic and end regular occupy similar regions as the permanently chaotic ones. This means that there is an annular region of initially chaotic particles that will migrate to constitute the bar, the majority of which will ultimately become regular. Finally, the orbits that begin regular and end chaotic can be traced back at $t=0$ to a less well-defined region outside the aforementioned annulus, but not extending as far as the outer disc. They also contribute to the bar, but their most noticeable feature is that these are the only kinds of orbits that significantly depart vertically from the $z=0$ plane, populating the so-called peanut, when the bar is viewed edge-on.

\end{enumerate}

Apart from the study of the disc, we also investigated the presence of chaotic and regular motion in the dark matter halo. For this, we similarly used an ensemble of halo orbits evolving in the presence of the time-dependent analytical model. Now we summarize the main results regarding the halo.

\begin{enumerate}

\item  We found that our analytical model was even able to recover the so-called halo bar, a prolate structure of the inner halo. This morphological feature is known to arise in self-consistent $N$-body simulations of strongly barred galaxies. Here, the halo potential itself remains spherically symmetric during the entire evolution, and yet the bar potential is able to induce the halo orbits to evolve into the shape of a halo bar. Like its $N$-body counterpart, the halo bar that developed in our ensemble of halo orbits is also less elongated and not as strong as the stellar bar itself. This reinforces the robustness of the analytical model, because the halo bar is quite a detailed morphological feature that had not been deliberately built into the model.

\item At any given time, chaotic motion is mainly expected to be found in the inner rather in the outer parts of the halo. Remarkably, the inner halo ($r<5$ kpc) is initially dominated by chaotic motion (as much as 70 percent chaotic). At later times, regular motion prevails at all radii, but it is still more present in the inner part. Also worthy of note is the fact that such high fraction of chaotic motion was not seen in any of the disc regions.

\item Globally, the time evolution of the fraction of chaotic motion in the halo is in the sense of decreasing monotonically, in any radial range. As with the disc, the tendency is for the halo orbits to become progressively more regular.

\end{enumerate}

% perspectives
Our analytical model has the advantage of offering a fully time-dependent and astrophysically realistic galaxy model, as indicated by the fact that it was successful in recovering several dynamical and morphological features of a barred galaxy. This work focused on one particular galaxy simulation, but the method may be extended to different galaxy types, taking as input the results of other $N$-body simulations. Here we focused on a strongly barred galaxy to maximise the effects we wished to explore. Clearly a natural development would be to compare the present results with alternative galaxy models of varying bar strengths, disc masses, halo profiles, etc. Such a systematic exploration would reveal to which parameters the dynamical stability is most sensitive. For example, given the high fraction of chaotic motion found in the inner halo, the question arises as to the role of the dark matter profile in determining that behaviour. One might explore whether a more cuspy inner halo would help or rather hinder the rise of regular motion. A further development would be the inclusion of models containing gas \citep{Patsis2009} and star formation. To this end, the hydrodynamical simulations of \cite*{AthMachRodMNRAS2013} would be ideally suited, since they already offer a systematic grid of models for galaxies with different halo triaxialities and different initial gas fractions, thus resulting in a variety of bar strength evolutions. More broadly, models derived from a fully cosmological hydrodynamical simulation of galaxy formation would offer an even more realistic scenario \citep[e.g][]{Valluri2013} than the usual models of isolated galaxies. Finally, a specific issues that merits further analysis is the behaviour of the X-shaped (or boxy/peanut) bulge \citep{Patsis2014a, Patsis2014b}, particularly in light of the recent interest in the kinematics and structure of the Milky Way's own bulge \citep{Saito2011,Zoccali2014}.

% end
The methods employed here have proven quite useful to the dynamical analyses of chaotic and regular motion, but this analytical approach is potentially suitable to various other applications in orbital studies.

%%%%%%%%%%%%%%%%%%%%%%%%%%%%%%%%%%%%%%%%%%%%%%%%%%%%%
\section*{Acknowledgments}

We thank the referee for valuable suggestions, which helped improve the paper. This work has made use of the computing facilities of the Laboratory of Astroinformatics (IAG/USP, NAT/Unicsul), whose purchase was made possible by the Brazilian agency FAPESP (grant 2009/54006-4) and the INCT-A. We also acknowledge support from \textit{Ci\^encia sem Fronteiras} (CNPq, Brazil) and from the Institut Henri Poincar\'e for its support through the \textit{Research in Paris} program.

%%%%%%%%%%%%%%%%%%%%%%%%%%%%%%%%%%%%%%%%%%%%%%%%%%%%%
\bibliographystyle{mnras}
\bibliography{MachMan}

\begin{thebibliography}{}
\makeatletter
\relax
\def\mn@urlcharsother{\let\do\@makeother \do\$\do\&\do\#\do\^\do\_\do\%\do\~}
\def\mn@doi{\begingroup\mn@urlcharsother \@ifnextchar [ {\mn@doi@}
  {\mn@doi@[]}}
\def\mn@doi@[#1]#2{\def\@tempa{#1}\ifx\@tempa\@empty \href
  {http://dx.doi.org/#2} {doi:#2}\else \href {http://dx.doi.org/#2} {#1}\fi
  \endgroup}
\def\mn@eprint#1#2{\mn@eprint@#1:#2::\@nil}
\def\mn@eprint@arXiv#1{\href {http://arxiv.org/abs/#1} {{\tt arXiv:#1}}}
\def\mn@eprint@dblp#1{\href {http://dblp.uni-trier.de/rec/bibtex/#1.xml}
  {dblp:#1}}
\def\mn@eprint@#1:#2:#3:#4\@nil{\def\@tempa {#1}\def\@tempb {#2}\def\@tempc
  {#3}\ifx \@tempc \@empty \let \@tempc \@tempb \let \@tempb \@tempa \fi \ifx
  \@tempb \@empty \def\@tempb {arXiv}\fi \@ifundefined
  {mn@eprint@\@tempb}{\@tempb:\@tempc}{\expandafter \expandafter \csname
  mn@eprint@\@tempb\endcsname \expandafter{\@tempc}}}

\bibitem[\protect\citeauthoryear{{Athanassoula}}{{Athanassoula}}{2005}]{Athanassoula2005}
{Athanassoula} E.,  2005, \mn@doi [\mnras] {10.1111/j.1365-2966.2005.08872.x},
  \href {http://adsabs.harvard.edu/abs/2005MNRAS.358.1477A} {358, 1477}

\bibitem[\protect\citeauthoryear{{Athanassoula}}{{Athanassoula}}{2007}]{Athanassoula2007}
{Athanassoula} E.,  2007, \mn@doi [\mnras] {10.1111/j.1365-2966.2007.11711.x},
  \href {http://adsabs.harvard.edu/abs/2007MNRAS.377.1569A} {377, 1569}

\bibitem[\protect\citeauthoryear{{Athanassoula}, {Romero-G{\'o}mez}  \&
  {Masdemont}}{{Athanassoula} et~al.}{2009a}]{AthaRomMas2009MNRAS}
{Athanassoula} E.,  {Romero-G{\'o}mez} M.,   {Masdemont} J.~J.,  2009a, \mn@doi
  [\mnras] {10.1111/j.1365-2966.2008.14273.x}, \href
  {http://adsabs.harvard.edu/abs/2009MNRAS.394...67A} {394, 67}

\bibitem[\protect\citeauthoryear{{Athanassoula}, {Romero-G{\'o}mez}, {Bosma}
  \& {Masdemont}}{{Athanassoula} et~al.}{2009b}]{AthRomBosMas2009MNRAS}
{Athanassoula} E.,  {Romero-G{\'o}mez} M.,  {Bosma} A.,   {Masdemont} J.~J.,
  2009b, \mn@doi [\mnras] {10.1111/j.1365-2966.2009.15583.x}, \href
  {http://adsabs.harvard.edu/abs/2009MNRAS.400.1706A} {400, 1706}

\bibitem[\protect\citeauthoryear{{Athanassoula}, {Romero-G{\'o}mez}, {Bosma}
  \& {Masdemont}}{{Athanassoula} et~al.}{2010}]{AthaRomBosMas2010MNRAS}
{Athanassoula} E.,  {Romero-G{\'o}mez} M.,  {Bosma} A.,   {Masdemont} J.~J.,
  2010, \mn@doi [\mnras] {10.1111/j.1365-2966.2010.17010.x}, \href
  {http://adsabs.harvard.edu/abs/2010MNRAS.407.1433A} {407, 1433}

\bibitem[\protect\citeauthoryear{{Athanassoula}, {Machado}  \&
  {Rodionov}}{{Athanassoula} et~al.}{2013}]{AthMachRodMNRAS2013}
{Athanassoula} E.,  {Machado} R.~E.~G.,   {Rodionov} S.~A.,  2013, \mn@doi
  [MNRAS] {10.1093/mnras/sts452}, \href
  {http://adsabs.harvard.edu/abs/2013MNRAS.429.1949A} {429, 1949}

\bibitem[\protect\citeauthoryear{{Benettin}, {Galgani}  \&
  {Strelcyn}}{{Benettin} et~al.}{1976}]{BenGalStr1976PRA}
{Benettin} G.,  {Galgani} L.,   {Strelcyn} J.-M.,  1976, \mn@doi [\pra]
  {10.1103/PhysRevA.14.2338}, \href
  {http://adsabs.harvard.edu/abs/1976PhRvA..14.2338B} {14, 2338}

\bibitem[\protect\citeauthoryear{{Benettin}, {Galgani}, {Giorgilli}  \&
  {Strelcyn}}{{Benettin} et~al.}{1980}]{Ben1980Mecc}
{Benettin} G.,  {Galgani} L.,  {Giorgilli} A.,   {Strelcyn} J.-M.,  1980,
  Meccanica, \href {http://adsabs.harvard.edu/abs/1980Mecc...15....9B} {15, 9}

\bibitem[\protect\citeauthoryear{{Bountis}, {Manos}  \&
  {Antonopoulos}}{{Bountis} et~al.}{2012}]{BouManAntCeMDA2012}
{Bountis} T.,  {Manos} T.,   {Antonopoulos} C.,  2012, \mn@doi [\CMDA]
  {10.1007/s10569-011-9392-9}, \href
  {http://adsabs.harvard.edu/abs/2012CeMDA.113...63B} {113, 63}

\bibitem[\protect\citeauthoryear{{Brunetti}, {Chiappini}  \&
  {Pfenniger}}{{Brunetti} et~al.}{2011}]{BruChiPfe2011A&A}
{Brunetti} M.,  {Chiappini} C.,   {Pfenniger} D.,  2011, \mn@doi [\aap]
  {10.1051/0004-6361/201117566}, \href
  {http://adsabs.harvard.edu/abs/2011A%26A...534A..75B} {534, A75}

\bibitem[\protect\citeauthoryear{{Col{\'{\i}}n}, {Valenzuela}  \&
  {Klypin}}{{Col{\'{\i}}n} et~al.}{2006}]{Colin2006}
{Col{\'{\i}}n} P.,  {Valenzuela} O.,   {Klypin} A.,  2006, \mn@doi [\apj]
  {10.1086/503791}, \href {http://adsabs.harvard.edu/abs/2006ApJ...644..687C}
  {644, 687}

\bibitem[\protect\citeauthoryear{Contopoulos}{Contopoulos}{2002}]{ContBook2002}
Contopoulos G.,  2002, Order and Chaos in Dynamical Astronomy.
Springer-Verlag, Berlin

\bibitem[\protect\citeauthoryear{{Contopoulos} \& {Harsoula}}{{Contopoulos} \&
  {Harsoula}}{2010}]{ConHar2010CeMDA}
{Contopoulos} G.,  {Harsoula} M.,  2010, \mn@doi [Celestial Mechanics and
  Dynamical Astronomy] {10.1007/s10569-010-9282-6}, \href
  {http://adsabs.harvard.edu/abs/2010CeMDA.107...77C} {107, 77}

\bibitem[\protect\citeauthoryear{{Contopoulos} \& {Harsoula}}{{Contopoulos} \&
  {Harsoula}}{2013}]{ConHar2013MNRAS}
{Contopoulos} G.,  {Harsoula} M.,  2013, \mn@doi [\mnras]
  {10.1093/mnras/stt1640}, \href
  {http://adsabs.harvard.edu/abs/2013MNRAS.436.1201C} {436, 1201}

\bibitem[\protect\citeauthoryear{{Contopoulos}, {Galgani}  \&
  {Giorgilli}}{{Contopoulos} et~al.}{1978}]{ConGalGio1978PRA}
{Contopoulos} G.,  {Galgani} L.,   {Giorgilli} A.,  1978, \mn@doi [\pra]
  {10.1103/PhysRevA.18.1183}, \href
  {http://adsabs.harvard.edu/abs/1978PhRvA..18.1183C} {18, 1183}

\bibitem[\protect\citeauthoryear{{Dehnen}}{{Dehnen}}{1993}]{DehnenMNRAS1993}
{Dehnen} W.,  1993, MNRAS, \href
  {http://adsabs.harvard.edu/abs/1993MNRAS.265..250D} {265, 250}

\bibitem[\protect\citeauthoryear{{Ferrers}}{{Ferrers}}{1877}]{Fer1877}
{Ferrers} N.~M.,  1877, \QJPAM, 14, 1

\bibitem[\protect\citeauthoryear{{Harsoula} \& {Kalapotharakos}}{{Harsoula} \&
  {Kalapotharakos}}{2009}]{HarKalMNRAS2009}
{Harsoula} M.,  {Kalapotharakos} C.,  2009, \mn@doi [MNRAS]
  {10.1111/j.1365-2966.2009.14427.x}, \href
  {http://adsabs.harvard.edu/abs/2009MNRAS.394.1605H} {394, 1605}

\bibitem[\protect\citeauthoryear{{Harsoula}, {Kalapotharakos}  \&
  {Contopoulos}}{{Harsoula} et~al.}{2011a}]{HarKalCont2011IJBC}
{Harsoula} M.,  {Kalapotharakos} C.,   {Contopoulos} G.,  2011a, \mn@doi
  [\IJBC] {10.1142/S0218127411029732}, \href
  {http://adsabs.harvard.edu/abs/2011IJBC...21.2221H} {21, 2221}

\bibitem[\protect\citeauthoryear{{Harsoula}, {Kalapotharakos}  \&
  {Contopoulos}}{{Harsoula} et~al.}{2011b}]{HarKalCont2011MNRAS}
{Harsoula} M.,  {Kalapotharakos} C.,   {Contopoulos} G.,  2011b, \mn@doi
  [\mnras] {10.1111/j.1365-2966.2010.17748.x}, \href
  {http://adsabs.harvard.edu/abs/2011MNRAS.411.1111H} {411, 1111}

\bibitem[\protect\citeauthoryear{{Hernquist}}{{Hernquist}}{1993}]{Hernquist1993}
{Hernquist} L.,  1993, \mn@doi [\apjs] {10.1086/191784}, \href
  {http://adsabs.harvard.edu/abs/1993ApJS...86..389H} {86, 389}

\bibitem[\protect\citeauthoryear{{Kandrup}, {Pogorelov}  \&
  {Sideris}}{{Kandrup} et~al.}{2000}]{Kandrupetal2000}
{Kandrup} H.~E.,  {Pogorelov} I.~V.,   {Sideris} I.~V.,  2000, \mn@doi [\mnras]
  {10.1046/j.1365-8711.2000.03097.x}, \href
  {http://adsabs.harvard.edu/abs/2000MNRAS.311..719K} {311, 719}

\bibitem[\protect\citeauthoryear{{Kaufmann} \& {Contopoulos}}{{Kaufmann} \&
  {Contopoulos}}{1996}]{KauCont1996A&A}
{Kaufmann} D.~E.,  {Contopoulos} G.,  1996, \aap, \href
  {http://adsabs.harvard.edu/abs/1996A%26A...309..381K} {309, 381}

\bibitem[\protect\citeauthoryear{Lichtenberg \& Lieberman}{Lichtenberg \&
  Lieberman}{1992}]{LichLieb}
Lichtenberg A.~J.,  Lieberman M.~A.,  1992, Regular and Chaotic Dynamics, 2nd
  edn.
No.~38 in Applied Mathematical Sciences, Springer-Verlag, New York, NY

\bibitem[\protect\citeauthoryear{{Machado} \& {Athanassoula}}{{Machado} \&
  {Athanassoula}}{2010}]{MachadoAthanassoula2010}
{Machado} R.~E.~G.,  {Athanassoula} E.,  2010, \mn@doi [MNRAS]
  {10.1111/j.1365-2966.2010.16890.x}, \href
  {http://adsabs.harvard.edu/abs/2010MNRAS.406.2386M} {406, 2386}

\bibitem[\protect\citeauthoryear{{Maffione}, {Darriba}, {Cincotta}  \&
  {Giordano}}{{Maffione} et~al.}{2011}]{MafDarCinGio2011CeMDA}
{Maffione} N.~P.,  {Darriba} L.~A.,  {Cincotta} P.~M.,   {Giordano} C.~M.,
  2011, \mn@doi [Celestial Mechanics and Dynamical Astronomy]
  {10.1007/s10569-011-9373-z}, \href
  {http://adsabs.harvard.edu/abs/2011CeMDA.111..285M} {111, 285}

\bibitem[\protect\citeauthoryear{{Maffione}, {Darriba}, {Cincotta}  \&
  {Giordano}}{{Maffione} et~al.}{2013}]{MafDarCinGio2013MNRAS}
{Maffione} N.~P.,  {Darriba} L.~A.,  {Cincotta} P.~M.,   {Giordano} C.~M.,
  2013, \mn@doi [\mnras] {10.1093/mnras/sts539}, \href
  {http://adsabs.harvard.edu/abs/2013MNRAS.429.2700M} {429, 2700}

\bibitem[\protect\citeauthoryear{{Maffione}, {G{\'o}mez}, {Cincotta},
  {Giordano}, {Cooper}  \& {O'Shea}}{{Maffione} et~al.}{2015}]{Maffione2015}
{Maffione} N.~P.,  {G{\'o}mez} F.~A.,  {Cincotta} P.~M.,  {Giordano} C.~M.,
  {Cooper} A.~P.,   {O'Shea} B.~W.,  2015, \mn@doi [\mnras]
  {10.1093/mnras/stv1778}, \href
  {http://adsabs.harvard.edu/abs/2015MNRAS.453.2830M} {453, 2830}

\bibitem[\protect\citeauthoryear{{Manos} \& {Athanassoula}}{{Manos} \&
  {Athanassoula}}{2011}]{ManAthMNRAS2011}
{Manos} T.,  {Athanassoula} E.,  2011, \mn@doi [MNRAS]
  {10.1111/j.1365-2966.2011.18734.x}, \href
  {http://adsabs.harvard.edu/abs/2011MNRAS.415..629M} {415, 629}

\bibitem[\protect\citeauthoryear{{Manos} \& {Machado}}{{Manos} \&
  {Machado}}{2014}]{ManosMachado2014}
{Manos} T.,  {Machado} R.~E.~G.,  2014, \mn@doi [\mnras]
  {10.1093/mnras/stt2355}, \href
  {http://adsabs.harvard.edu/abs/2014MNRAS.438.2201M} {438, 2201}

\bibitem[\protect\citeauthoryear{{Manos}, {Skokos}  \& Antonopoulos}{{Manos}
  et~al.}{2012}]{ManSkoAnt}
{Manos} T.,  {Skokos} C.,   Antonopoulos C.,  2012, \mn@doi [\IJBC]
  {10.1142/S0218127412502185}, 22, 1250218

\bibitem[\protect\citeauthoryear{{Manos}, {Bountis}  \& {Skokos}}{{Manos}
  et~al.}{2013}]{ManBouSkoJPhA2013}
{Manos} T.,  {Bountis} T.,   {Skokos} C.,  2013, \mn@doi [\JPhysAmt]
  {10.1088/1751-8113/46/25/254017}, \href
  {http://adsabs.harvard.edu/abs/2013JPhA...46y4017M} {46, 254017}

\bibitem[\protect\citeauthoryear{{Miyamoto} \& {Nagai}}{{Miyamoto} \&
  {Nagai}}{1975}]{MNPASJ1975}
{Miyamoto} M.,  {Nagai} R.,  1975, PASJ, \href
  {http://adsabs.harvard.edu/abs/1975PASJ...27..533M} {27, 533}

\bibitem[\protect\citeauthoryear{{Patsis}}{{Patsis}}{2006}]{Pat2006MNRAS}
{Patsis} P.~A.,  2006, \mn@doi [\mnras] {10.1111/j.1745-3933.2006.00174.x},
  \href {http://adsabs.harvard.edu/abs/2006MNRAS.369L..56P} {369, L56}

\bibitem[\protect\citeauthoryear{{Patsis} \& {Katsanikas}}{{Patsis} \&
  {Katsanikas}}{2014a}]{Patsis2014a}
{Patsis} P.~A.,  {Katsanikas} M.,  2014a, \mn@doi [\mnras]
  {10.1093/mnras/stu1988}, \href
  {http://adsabs.harvard.edu/abs/2014MNRAS.445.3525P} {445, 3525}

\bibitem[\protect\citeauthoryear{{Patsis} \& {Katsanikas}}{{Patsis} \&
  {Katsanikas}}{2014b}]{Patsis2014b}
{Patsis} P.~A.,  {Katsanikas} M.,  2014b, \mn@doi [\mnras]
  {10.1093/mnras/stu1970}, \href
  {http://adsabs.harvard.edu/abs/2014MNRAS.445.3546P} {445, 3546}

\bibitem[\protect\citeauthoryear{{Patsis}, {Athanassoula}  \&
  {Quillen}}{{Patsis} et~al.}{1997}]{PatAthQui1997ApJ}
{Patsis} P.~A.,  {Athanassoula} E.,   {Quillen} A.~C.,  1997, \mn@doi [\apj]
  {10.1086/304287}, \href {http://adsabs.harvard.edu/abs/1997ApJ...483..731P}
  {483, 731}

\bibitem[\protect\citeauthoryear{{Patsis}, {Kaufmann}, {Gottesman}  \&
  {Boonyasait}}{{Patsis} et~al.}{2009}]{Patsis2009}
{Patsis} P.~A.,  {Kaufmann} D.~E.,  {Gottesman} S.~T.,   {Boonyasait} V.,
  2009, \mn@doi [\mnras] {10.1111/j.1365-2966.2008.14335.x}, \href
  {http://adsabs.harvard.edu/abs/2009MNRAS.394..142P} {394, 142}

\bibitem[\protect\citeauthoryear{{Pfenniger}}{{Pfenniger}}{1984}]{PfeA&A1984a}
{Pfenniger} D.,  1984, A\&A, \href
  {http://adsabs.harvard.edu/abs/1984A%26A...134..373P} {134, 373}

\bibitem[\protect\citeauthoryear{{Price-Whelan}, {Johnston}, {Valluri},
  {Pearson}, {K{\"u}pper}  \& {Hogg}}{{Price-Whelan}
  et~al.}{2016}]{Price-Whelan2016}
{Price-Whelan} A.~M.,  {Johnston} K.~V.,  {Valluri} M.,  {Pearson} S.,
  {K{\"u}pper} A.~H.~W.,   {Hogg} D.~W.,  2016, \mn@doi [\mnras]
  {10.1093/mnras/stv2383}, \href
  {http://adsabs.harvard.edu/abs/2016MNRAS.455.1079P} {455, 1079}

\bibitem[\protect\citeauthoryear{{Romero-G{\'o}mez}, {Masdemont},
  {Athanassoula}  \& {Garc{\'{\i}}a-G{\'o}mez}}{{Romero-G{\'o}mez}
  et~al.}{2006}]{RomMasAthGar2006A&A}
{Romero-G{\'o}mez} M.,  {Masdemont} J.~J.,  {Athanassoula} E.,
  {Garc{\'{\i}}a-G{\'o}mez} C.,  2006, \mn@doi [\aap]
  {10.1051/0004-6361:20054653}, \href
  {http://adsabs.harvard.edu/abs/2006A%26A...453...39R} {453, 39}

\bibitem[\protect\citeauthoryear{{Romero-G{\'o}mez}, {Athanassoula},
  {Masdemont}  \& {Garc{\'{\i}}a-G{\'o}mez}}{{Romero-G{\'o}mez}
  et~al.}{2007}]{RomAthMasGar2007A&A}
{Romero-G{\'o}mez} M.,  {Athanassoula} E.,  {Masdemont} J.~J.,
  {Garc{\'{\i}}a-G{\'o}mez} C.,  2007, \mn@doi [\aap]
  {10.1051/0004-6361:20077504}, \href
  {http://adsabs.harvard.edu/abs/2007A%26A...472...63R} {472, 63}

\bibitem[\protect\citeauthoryear{{Saito}, {Zoccali}, {McWilliam}, {Minniti},
  {Gonzalez}  \& {Hill}}{{Saito} et~al.}{2011}]{Saito2011}
{Saito} R.~K.,  {Zoccali} M.,  {McWilliam} A.,  {Minniti} D.,  {Gonzalez}
  O.~A.,   {Hill} V.,  2011, \mn@doi [\aj] {10.1088/0004-6256/142/3/76}, \href
  {http://adsabs.harvard.edu/abs/2011AJ....142...76S} {142, 76}

\bibitem[\protect\citeauthoryear{{Skokos}}{{Skokos}}{2010}]{SkoLNP2010}
{Skokos} C.,  2010, \mn@doi [\LNP] {10.1007/978-3-642-04458-8_2}, \href
  {http://adsabs.harvard.edu/abs/2010LNP...790...63S} {790, 63}

\bibitem[\protect\citeauthoryear{{Skokos} \& {Manos}}{{Skokos} \&
  {Manos}}{2016}]{SkoManLNP2016}
{Skokos} C.,  {Manos} T.,  2016, \mn@doi [\LNP] {10.1007/978-3-662-48410-4},
  915, (in press), arXiv:1412.7401

\bibitem[\protect\citeauthoryear{{Skokos}, {Bountis}  \&
  {Antonopoulos}}{{Skokos} et~al.}{2007}]{SkoBouAntPhyD2007}
{Skokos} C.,  {Bountis} T.~C.,   {Antonopoulos} C.,  2007, \mn@doi [Physica D]
  {10.1016/j.physd.2007.04.004}, \href
  {http://adsabs.harvard.edu/abs/2007PhyD..231...30S} {231, 30}

\bibitem[\protect\citeauthoryear{{Skokos}, {Bountis}  \&
  {Antonopoulos}}{{Skokos} et~al.}{2008}]{SkoBouAntEPJST2008}
{Skokos} C.,  {Bountis} T.,   {Antonopoulos} C.,  2008, \mn@doi [\EPJST]
  {10.1140/epjst/e2008-00844-2}, \href
  {http://adsabs.harvard.edu/abs/2008EPJST.165....5S} {165, 5}

\bibitem[\protect\citeauthoryear{{Skokos}, {Gottwald}  \& {Laskar}}{{Skokos}
  et~al.}{2016}]{LNinP2016}
{Skokos} C.,  {Gottwald} G.~A.,   {Laskar} J.,  eds, 2016, Chaos Detection and
  Predictability.
~\LNP Vol. 915, Springer-Verlag, Berlin-Heidelberg,
  \mn@doi{10.1007/978-3-662-48410-4}

\bibitem[\protect\citeauthoryear{{Terzi{\'c}} \& {Kandrup}}{{Terzi{\'c}} \&
  {Kandrup}}{2004}]{TerKand2004MNRAS}
{Terzi{\'c}} B.,  {Kandrup} H.~E.,  2004, \mn@doi [\mnras]
  {10.1111/j.1365-2966.2004.07256.x}, \href
  {http://adsabs.harvard.edu/abs/2004MNRAS.347..957T} {347, 957}

\bibitem[\protect\citeauthoryear{{Tsoutsis}, {Kalapotharakos}, {Efthymiopoulos}
   \& {Contopoulos}}{{Tsoutsis} et~al.}{2009}]{TsoKalEftCon2009A&A}
{Tsoutsis} P.,  {Kalapotharakos} C.,  {Efthymiopoulos} C.,   {Contopoulos} G.,
  2009, \mn@doi [\aap] {10.1051/0004-6361:200810149}, \href
  {http://adsabs.harvard.edu/abs/2009A%26A...495..743T} {495, 743}

\bibitem[\protect\citeauthoryear{{Valluri}, {Debattista}, {Quinn}  \&
  {Moore}}{{Valluri} et~al.}{2010}]{Valluri2010}
{Valluri} M.,  {Debattista} V.~P.,  {Quinn} T.,   {Moore} B.,  2010, \mn@doi
  [\mnras] {10.1111/j.1365-2966.2009.16192.x}, \href
  {http://adsabs.harvard.edu/abs/2010MNRAS.403..525V} {403, 525}

\bibitem[\protect\citeauthoryear{{Valluri}, {Debattista}, {Quinn}, {Ro{\v
  s}kar}  \& {Wadsley}}{{Valluri} et~al.}{2012}]{Valluri2012}
{Valluri} M.,  {Debattista} V.~P.,  {Quinn} T.~R.,  {Ro{\v s}kar} R.,
  {Wadsley} J.,  2012, \mn@doi [\mnras] {10.1111/j.1365-2966.2011.19853.x},
  \href {http://adsabs.harvard.edu/abs/2012MNRAS.419.1951V} {419, 1951}

\bibitem[\protect\citeauthoryear{{Valluri}, {Debattista}, {Stinson}, {Bailin},
  {Quinn}, {Couchman}  \& {Wadsley}}{{Valluri} et~al.}{2013}]{Valluri2013}
{Valluri} M.,  {Debattista} V.~P.,  {Stinson} G.~S.,  {Bailin} J.,  {Quinn}
  T.~R.,  {Couchman} H.~M.~P.,   {Wadsley} J.,  2013, \mn@doi [\apj]
  {10.1088/0004-637X/767/1/93}, \href
  {http://adsabs.harvard.edu/abs/2013ApJ...767...93V} {767, 93}

\bibitem[\protect\citeauthoryear{{Vasiliev} \& {Athanassoula}}{{Vasiliev} \&
  {Athanassoula}}{2015}]{Vasiliev2015}
{Vasiliev} E.,  {Athanassoula} E.,  2015, \mn@doi [\mnras]
  {10.1093/mnras/stv805}, \href
  {http://adsabs.harvard.edu/abs/2015MNRAS.450.2842V} {450, 2842}

\bibitem[\protect\citeauthoryear{{Weinberg}}{{Weinberg}}{2015b}]{Weinberg2015a}
{Weinberg} M.~D.,  2015b, preprint, \href
  {http://adsabs.harvard.edu/abs/2015arXiv150806855W} {} (\mn@eprint {arXiv}
  {1508.06855})

\bibitem[\protect\citeauthoryear{{Weinberg}}{{Weinberg}}{2015a}]{Weinberg2015b}
{Weinberg} M.~D.,  2015a, preprint, \href
  {http://adsabs.harvard.edu/abs/2015arXiv150805959W} {} (\mn@eprint {arXiv}
  {1508.05959})

\bibitem[\protect\citeauthoryear{{Zoccali} et~al.,}{{Zoccali}
  et~al.}{2014}]{Zoccali2014}
{Zoccali} M.,  et~al., 2014, \mn@doi [\aap] {10.1051/0004-6361/201323120},
  \href {http://adsabs.harvard.edu/abs/2014A%26A...562A..66Z} {562, A66}

\makeatother
\end{thebibliography}
\bsp
\label{lastpage}
\end{document}